\documentclass{aa}
\usepackage{graphicx}
\usepackage{astrobib}
\usepackage{journals}
\begin{document}
   \title{The extreme flare in III~Zw~2:}

   \subtitle{Evolution of a radio jet in a Seyfert galaxy}

   \author{A. Brunthaler\inst{1,2},
	  H. Falcke\inst{3,4},
	  G.C. Bower\inst{5},
	  M.F. Aller\inst{6},
	  H.D. Aller\inst{6},
	  \and
	  H. Ter\"asranta\inst{7}
          }

   \offprints{brunthaler@jive.nl}

   \institute{Max-Planck-Institut f\"ur Radioastonomie, Auf dem H\"ugel 69,
              53121 Bonn, Germany
         \and
         Joint Institute for VLBI in Europe, Postbus 2, 7990 AA
         Dwingeloo, The Netherlands
			\and
			ASTRON, Postbus 2, 7990 AA Dwingeloo, The Netherlands
         \and
Department of Astrophysics, Radboud Universiteit Nijmegen, Postbus
              9010, 6500 GL Nijmegen, The Netherlands 			
			\and 
	      Radio Astronomy Laboratory, University of California at Berkeley,
	      601 Campbell Hall, CA 94720, USA
	 \and
              University of Michigan, Astronomy Department, Ann Arbor, 
	      MI 48109-1090, USA
	 \and
	      Mets\"ahovi Radio Observatory, Helsinki University of
              Technology, Metsahovintie 114, 02540 Kylm\"al\"a, Finland
             }

   \date{Received ; accepted }

   \abstract{ A very detailed monitoring of a radio flare in the Seyfert I
	  galaxy III~Zw~2 with the VLA and the VLBA is presented. The
	  relative astrometry in the VLBA observations was precise on a
	  level of a few $\mu$as. Spectral and spatial evolution of the
	  source are closely linked and these observations allowed us to
	  study in great detail a textbook example of a synchrotron
	  self-absorbed jet. We observe a phase where the jet gets
	  frustrated, without expansion and no spectral evolution.
     Then the jet breaks free and starts to expand with
	  apparent superluminal motion. This expansion is accompanied by a
	  strong spectral evolution.  The results are a good confirmation of
	  synchrotron theory and equipartition for jets.

   }
   \authorrunning{A. Brunthaler, H. Falcke, G.C. Bower et al.}
   \titlerunning{III~Zw~2: Evolution of a radio jet in a Seyfert galaxy}
   \maketitle
%

\section{Introduction}

The radio properties of quasars with otherwise very similar optical
properties can be markedly different. There is a clear dichotomy
between radio-loud and radio-quiet quasars in optically selected
samples. The radio-loudness is usually characterized by the
radio-to-optical flux ratio. In the PG quasar sample, which is
probably the best studied quasar sample in the radio and optical
(\citeNP{KellermannSramekSchmidt1989}; \citeNP{BorosonGreen1992}),
radio-loud and radio-weak quasars separate cleanly in two distinct
populations (e.g.~\citeNP{KellermannSramekSchmidt1989}).

It is known that radio-loud AGN almost never reside in late type,
i.e.~spiral galaxies (e.g.~\citeNP{KirhakosBahcallSchneider1999};
\citeNP{BahcallKirhakosSchneider1995}) whereas radio-quiet quasars
appear both in spiral and in elliptical host galaxies.  Furthermore,
all relativistically boosted jets with superluminal motion and typical
blazars have been detected in early type galaxies
(e.g.~\citeNP{ScarpaUrryFalomo2000}). It is still unclear, why AGN in
spiral galaxies, at the same optical luminosity as their elliptical
counterparts, should not be able to produce the powerful, relativistic
jets seen in radio galaxies.

However, a few sources with intermediate radio-to-optical ratios
appear to be neither radio-loud nor radio-quiet.  They form a distinct
subclass with very similar radio morphological and spectral
properties.  They all have a compact core at arcsecond scales and a flat and
variable spectrum in common. These properties are very similar to the
ones of radio cores in radio-loud quasars, but their low
radio-to-optical ratio and their low extended steep-spectrum emission
is atypical for radio-loud quasars.~\citeN{MillerRawlingsSaunders1993}
and \citeN{FalckeSherwoodPatnaik1996} have identified a number of these
sources, called ``radio-intermediate quasars'' (RIQs), and suggested
that they might be relativistically boosted radio-weak quasars or
``radio-weak blazars''. This would imply that most, if not all,
radio-quiet quasars also have relativistic jets. In fact, Very Long
Baseline Interferometry (VLBI) observations of radio-quiet quasars
already have shown high-brightness temperature radio cores and jets
(\citeNP{FalckePatnaikSherwood1996}; \citeNP{BlundellBeasley1998}). A
crucial test of the relativistic jet hypothesis is the search for
apparent superluminal motion in these sources. A prime candidate for
detecting this is the brightest radio source in the RIQ sample,
III~Zw~2, which we discuss in this paper. 

III~Zw~2 (PG 0007+106, Mrk 1501, $z=0.089$) was discovered by
\citeN{Zwicky1967}, classified as a Seyfert I galaxy (e.g., \citeNP{Arp1968};
\citeNP{KhachikianWeedman1974}; \citeNP{Osterbrock1977}), and later also
included in the PG quasar sample (\citeNP{SchmidtGreen1983}). The host galaxy
was classified as a spiral (e.g. \citeNP{HutchingsCampbell1983}) and a
disk model was later confirmed by fitting of model isophotes to near-IR images
(\citeNP{TaylorDunlopHughes1996}). A spiral arm was claimed
(\citeNP{Hutchings1983}) but recent observations suggest a tidal arm
with several knots of star forming regions (\citeNP{SuraceSandersEvans2001}).
III~Zw~2 is the brightest member of a group of galaxies and an extended low
surface brightness emission surrounding all the galaxies suggests that
there are perhaps interactions between the galaxies
(\citeNP{SuraceSandersEvans2001}).

The source has shown extreme variability at radio wavelengths with at least
20-fold increases in radio flux density within 4 years
(\citeNP{AllerAllerLatimer1985}). III~Zw~2 is also known to be variable in the
optical (\citeNP{Lloyd1984}; \citeNP{ClementsSmithAller1995}) and X-ray
(\citeNP{KaastradeKorte1988}). \citeN{SalviPageStevens2002} compare the
long term radio light curves of III~Zw~2 with optical, IR and X-ray
light curves and find indications for correlated flux variations from
radio to X-ray wavelengths. Unfortunately, the time sampling at other
wavelengths than radio is very poor.

III~Zw~2 is a core-dominated flat-spectrum AGN with only a faint
extended structure (see~\citeNP{UngerLawrenceWilson1987} and
Section~\ref{iiizw2_ext}). The weak extended radio emission and the
host galaxy is quite typical for a Seyfert galaxy. Its [O{\sc III}]
luminosity is a mere factor three brighter than that of a bright
Seyfert galaxy like Mrk~3
(e.g.~\citeNP{Alonso-HerreroWardKotilainen1997}) which explains why it
has been classified as either a Seyfert galaxy or a quasar. In this
luminosity region a distinction between the two may not be of much
significance.

Earlier VLBI observations of the source have only shown a
high-brightness temperature unresolved core
(\citeNP{FalckeSherwoodPatnaik1996},
~\citeNP{KellermannVermeulenZensus1998}) and Millimeter-VLBI
observations by~\citeN{FalckeBowerLobanov1999} just barely resolved the
source into two very compact components. A broadband radio spectrum
showed a highly peaked spectrum which was well explained by a very
compact source and synchrotron self-absorption.  

The unique and simple structure and timescales of radio outbursts within
5 years makes III~Zw~2 and ideal source to study radio-jet evolution
relevant also to radio galaxies.

In section 2, we will describe our Very Large Array (VLA) and Very Long
Baseline Array (VLBA) observations. before results from VLA monitoring
are presented in sections 3.1 -- 3.3. We then describe the results from
VLBI observations in section 3.4. In section 4 we will discuss the
results.

\section{Observations}

\begin{table}
      \caption[]{Total observing time $t_{obs}$, fraction of observing time
on III~Zw~2 at 43 GHz $f_{43}$ and 15 GHz $f_{15}$ and fraction of observing 
time on phase-reference quasar $f_{quasar}$ for the VLBA observations.}
         \label{vlbainfo}
      \[
  \begin{tabular}{p{0.18\linewidth}|cp{0.14\linewidth}cp{0.14\linewidth}cp{0.14\linewidth}cp{0.14\linewidth}}
           \hline
 Date & $t_{obs}$  & $f_{43}$ &$f_{15}$& $f_{quasar}$  \\
            \hline
           1998/02/16 & 8 h  &0.75 &0.25 &--\\
           1998/06/13 & 9 h  &0.75 &0.25 &--\\
           1998/09/14 & 8 h  &0.75 &0.25 &--\\
           1998/12/12 & 8 h  &0.75 &0.25 &--\\
           1999/07/15 & 6 h  &0.75 &0.25 &--\\
	   1999/11/12 & 8 h  &0.33 &0.33 &0.33\\
	   2000/07/22 & 6 h  &0.33 &0.33 &0.33\\
	   2000/08/27 & 8 h  &0.33 &0.33 &0.33\\
	   2000/09/06 & 8 h  &0.33 &0.33 &0.33\\
            \hline
         \end{tabular}
      \]
   \end{table}

In 1996 III~Zw~2 started a new major radio outburst and we initiated a
target of opportunity program to monitor the spectral evolution of the
burst with the VLA and its structural evolution with the VLBA with
excellent relative astrometry of the component separation.

We observed III~Zw~2 with the VLA 41 times from 1998 September until
September 2001 in intervals of roughly one month. The observations were
made at six frequencies ranging from 1.4 GHz to 43 GHz. Results from an
observation on 1998 May 21 at 350 MHz and on 1999 July 7 at 327.5 MHz are
also presented here.
The source 3C48 was used as the primary flux density calibrator, and
III~Zw~2 was self-calibrated and mapped with the Astronomical Image Processing
System (AIPS).

We also used the monitoring data at 8 and 15 GHz obtained with the
Michigan 26 m telescope, and at 22 and 37 GHz from the Mets\"ahovi
radio telescope. The single dish data are important for placing the VLA
and VLBA data in context, as they include a larger time window and are
more closely spaced.


We observed III~Zw~2 with the VLBA nine times over a period of 2.5 years at
15 and 43 GHz.
Details of this observations are given in Table~\ref{vlbainfo}. In the last
four epochs we included the background quasar J0011+0823 at 15 GHz as
phase-reference source. For the second epoch, we used the Effelsberg 100 m
telescope in combination
with the VLBA. We observed four 8 MHz bands, each at right and left circular
polarization. The initial calibration was performed with the AIPS package.
A-priori amplitude calibration was applied using system temperature
measurements and standard gain curves. Fringes were found in the III~Zw~2
data on all baselines. The data were self-calibrated and mapped using the
software package DIFMAP (\citeNP{ShepherdPearsonTaylor1994}). We started with
phase-only self-calibration and later included phase-amplitude self-calibration
with solution intervals slowly decreasing down to one minute. Results
of the first five VLBA observations at 43 GHz were reported
by~\citeN{BrunthalerFalckeBower2000}.

\section{Results}
\subsection{Extended emission of III~Zw~2}
\label{iiizw2_ext}

\begin{figure}
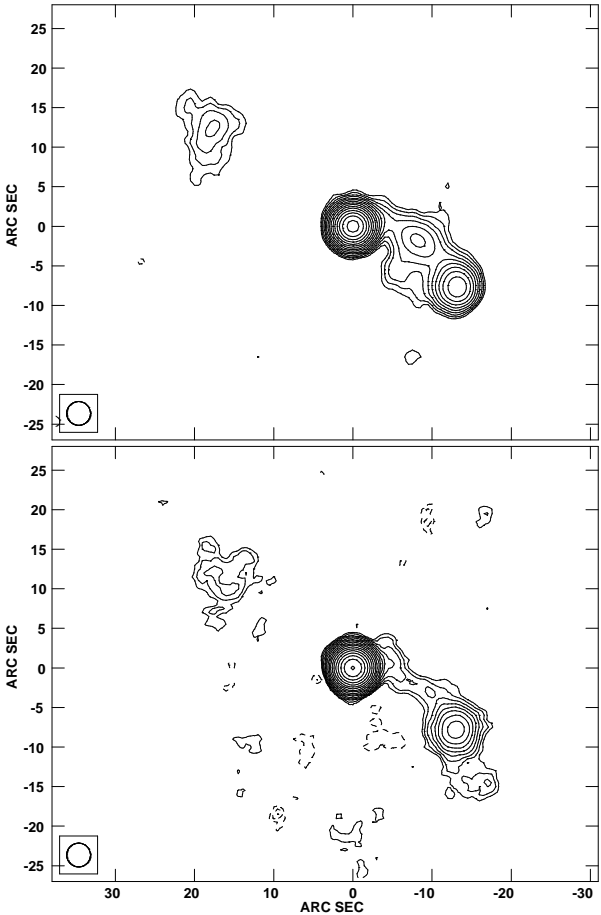

\resizebox{\hsize}{!}{\includegraphics[bbllx=2.5cm,bburx=16.7cm,bblly=3.8cm,bbury=25.2cm,clip=,angle=-90]{iiizw2_l.ps}}
\resizebox{\hsize}{!}{\includegraphics[bbllx=2.5cm,bburx=17.7cm,bblly=3.8cm,bbury=25.2cm,clip=,angle=-90]{iiizw2_c.ps}}
\caption{Combined VLA map of 11 epochs (A,B,C and D-array) of III~Zw~2 at 
1.4 GHz (top) and 4.8 GHz (bottom). All maps were convolved with a beam
of 2 $\times$2 arcseconds to detect faint extended emission.  The
contours start at 0.26 mJy and 0.15 mJy at 1.4 GHz and 4.8 GHz
respectively and increase with a factor of $\sqrt 2$.}
\label{lc-map}
\end{figure}

\begin{figure}
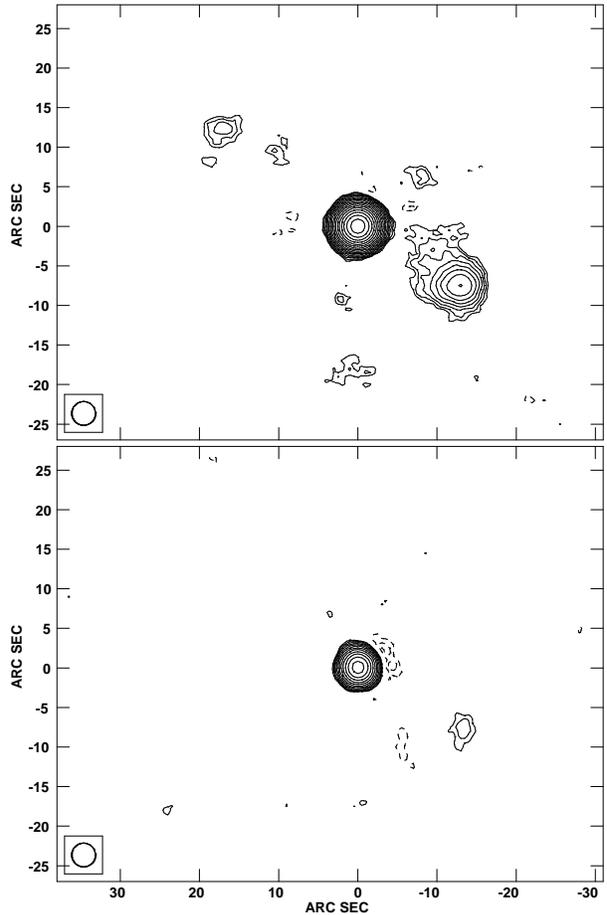

\resizebox{\hsize}{!}{\includegraphics[bbllx=2.5cm,bburx=16.7cm,bblly=3.8cm,bbury=25.2cm,clip=,angle=-90]{iiizw2_x.ps}}
\resizebox{\hsize}{!}{\includegraphics[bbllx=2.5cm,bburx=17.7cm,bblly=3.8cm,bbury=25.2cm,clip=,angle=-90]{iiizw2_u.ps}}
\caption{Combined VLA map of 11 epochs (A,B,C and D-array) of III~Zw~2 at 
8.4 GHz (top) and 15 GHz (bottom). All maps were convolved 
with a  beam of 2$\times$2 arcseconds to detect faint extended
emission. The contours start at 0.12 mJy and 0.6 mJy at 8.4 GHz and 15 GHz
respectively and increase with a factor of $\sqrt 2$.}
\label{xu-map}
\end{figure}

\citeN{UngerLawrenceWilson1987} discovered a weak radio component
$15.4''$ (23 kpc, with an angular size distance of $d_A\sim307.4$ Mpc;
$H_0=75~km~sec^{-1}~Mpc^{-1}$, $q_0=0.5$ as used in this paper)
southwest of the nucleus. This detection was confirmed later
(\citeNP{KukulaDunlopHughes1998}; \citeNP{FalckeBowerLobanov1999}), but
no additional extended radio emission was found.

To study the extended structure in more detail, we combined the raw
data of eleven VLA observations. In the combined data we used data from the
VLA in A, B, C and D configuration. Since the nucleus is highly variable, we
subtracted it from the uv-data before combining the data. The combined data
set was then self-calibrated and mapped.
The combined VLA maps at 1.4, 4.8, 8.4 and 15 GHz are shown in Fig.
\ref{lc-map} and \ref{xu-map}. The 4.8, 8.4 and 15 GHz maps were convolved
with a large beam of 2$\times$2 arcseconds to detect faint extended
structure. We detected the southwestern component at all four frequencies.
This radio lobe or hotspot is connected to the nucleus with a jet-like
structure visible at 1.4, 4.8 and 8.4 GHz. At 1.4 GHz one sees an indication
that the jet is ejected in northwestern direction and gets deflected by almost
$90^{\circ}$ towards the southwestern lobe. This is also in accordance with the
direction of the jet on sub-parsec scales (see section~\ref{vlbi}).


We also discovered a weaker secondary radio lobe $21.9''$ (32.6 kpc) on the
opposite side of the galaxy at 1.4 to 8.4 GHz. If one assumes equal expansion
velocities for both lobes,
a simple time travel argument (e.g., \citeNP{RyleLongair1967}) would suggest
that the weaker northeastern lobe is approaching and the brighter southwestern
lobe is receding. However, this scenario can not explain the differences in
flux density between the two lobes. One would expect the approaching lobe to be
brighter due to relativistic boosting of the emission.

Hence it is more likely that the armlength difference is explained by an
asymmetric expansion of the the two lobes due to different intrinsic
velocities or differences in the ambient medium, i.e. the medium in the
southwest of III~Zw~2 has a higher density than the medium in the
northeast. This is supported by the fact that there is a close
companion galaxy only $\sim 30''$ to the south.

\begin{figure}
\resizebox{\hsize}{!}{\includegraphics[bbllx=9.2cm,bburx=19.6cm,bblly=1.9cm,bbury=16.6cm,clip=,angle=-90]{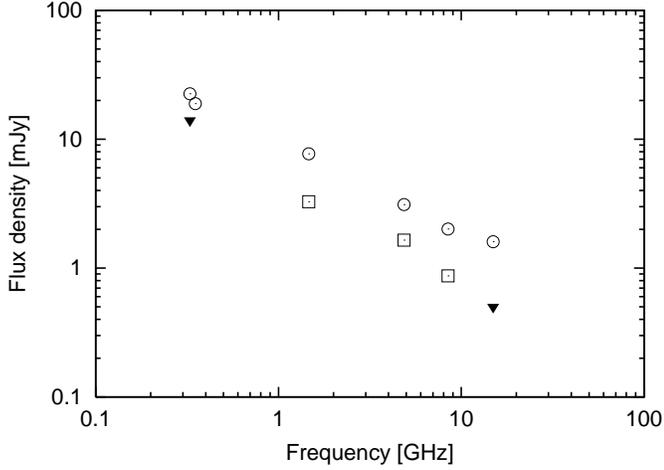}}
\caption{Spectra of the southwestern (circles) and northeastern (squares) 
radio lobes. The black triangles are upper limits for the northeastern lobe.}
\label{lobes}
\end{figure}

The spectra
of the two radio lobes are shown in Fig~\ref{lobes}. The 350 and 327.5 MHz
data were from our observations on 1998 May 21 and 1999 July 7 respectively
where we also detected the southwestern component. Both radio lobes have a
steep spectrum with spectral indices of $\alpha =
-0.57$~to~$-1.15$. Values around -0.7 are typical for synchrotron
emission of optically thin radio lobes of radio galaxies. There is no
break or a steepening of the spectrum towards higher frequencies. This
indicates that both radio lobes are still active and powered by the
central engine. Otherwise, the high energy electrons would have lost
most of their energy due to radiation losses. This would lead to a
steepening in the spectrum at higher frequencies.

\subsection{Variability}

\begin{figure}
\resizebox{\hsize}{!}{\includegraphics[bbllx=9.2cm,bburx=18.2cm,bblly=1.9cm,bbury=16.6cm,clip=,angle=-90]{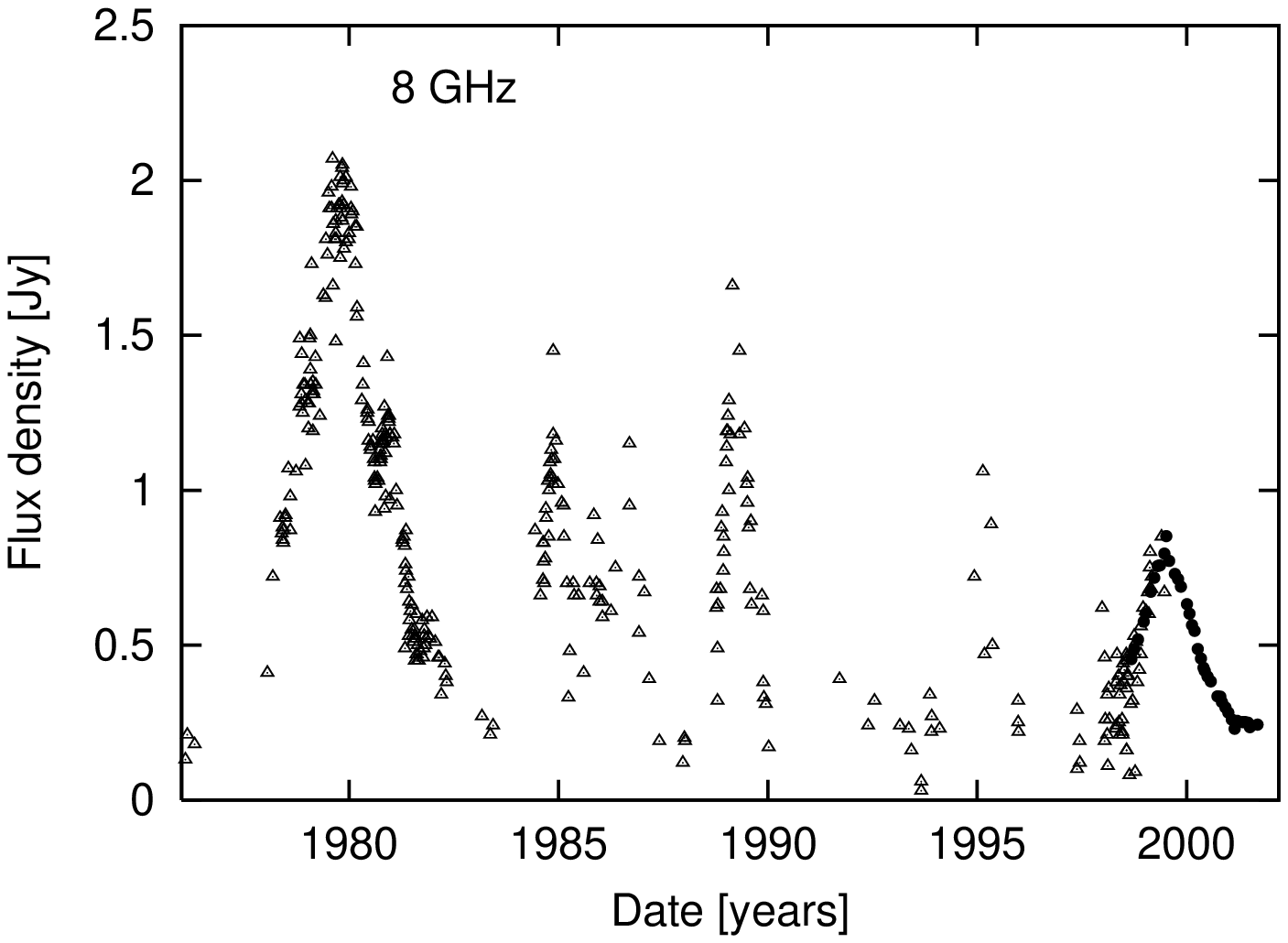}}
\resizebox{\hsize}{!}{\includegraphics[bbllx=9.2cm,bburx=18.2cm,bblly=1.9cm,bbury=16.6cm,clip=,angle=-90]{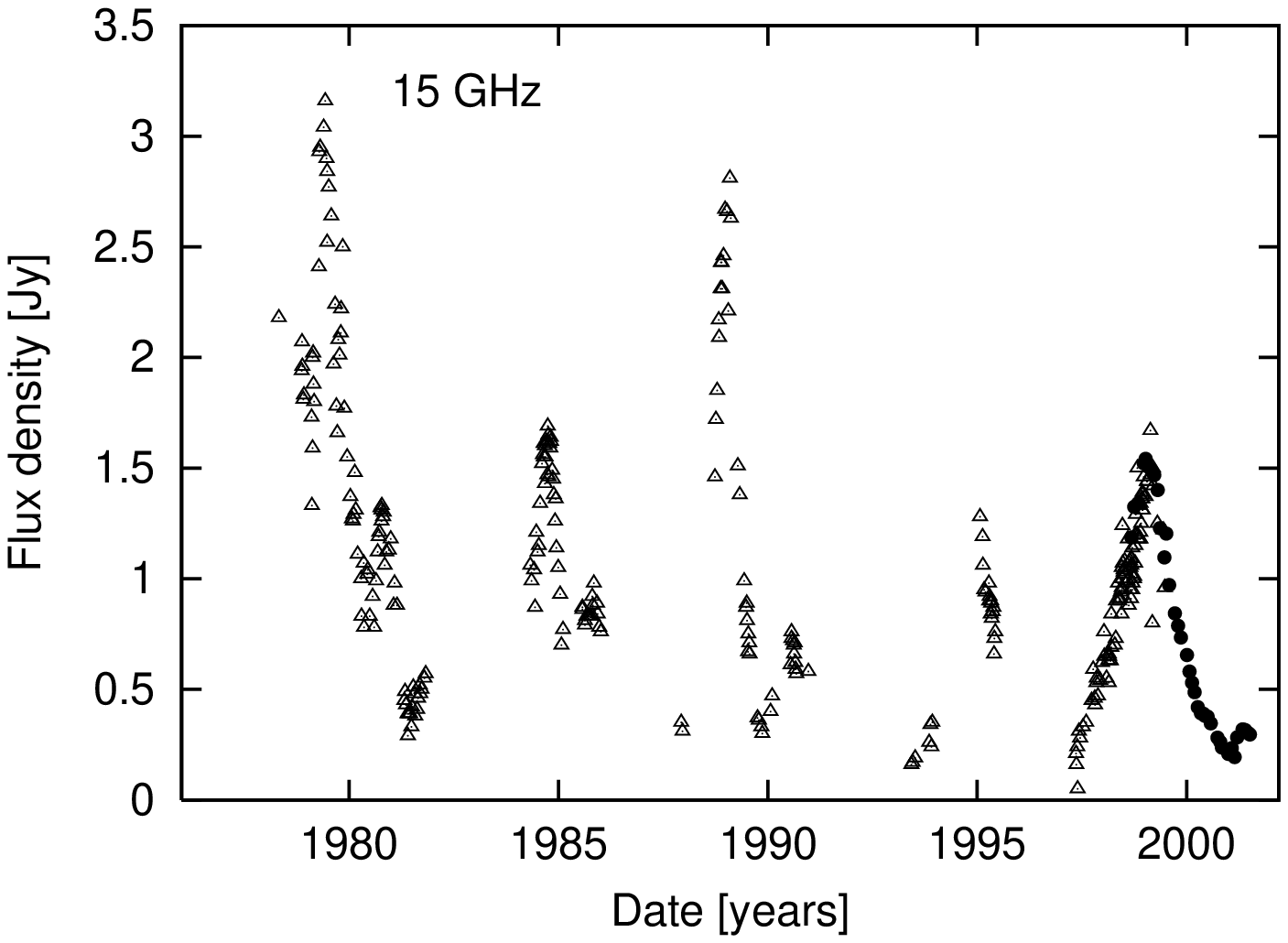}}
\resizebox{\hsize}{!}{\includegraphics[bbllx=9.2cm,bburx=18.2cm,bblly=1.9cm,bbury=16.6cm,clip=,angle=-90]{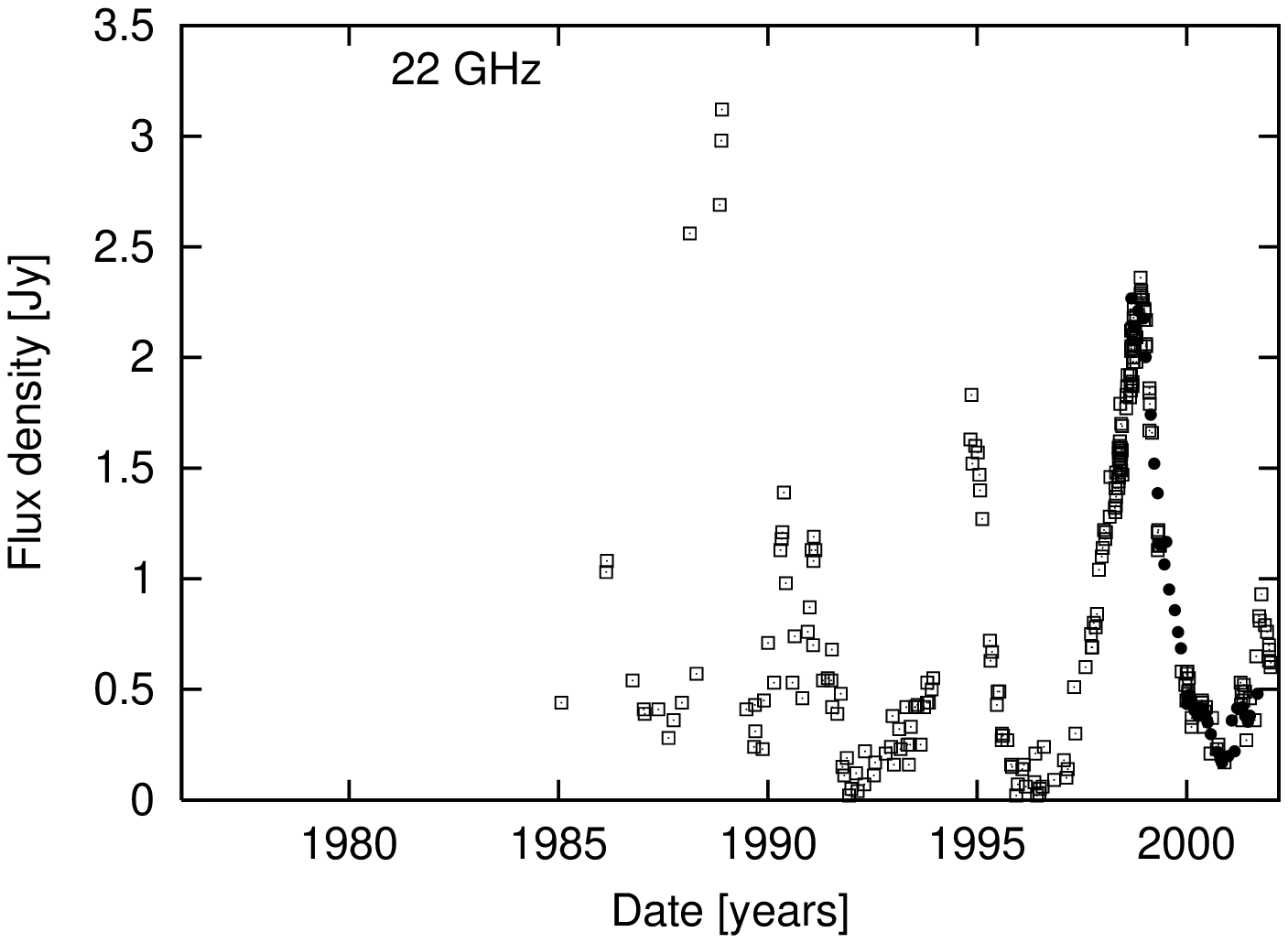}}
\resizebox{\hsize}{!}{\includegraphics[bbllx=9.2cm,bburx=19.6cm,bblly=1.9cm,bbury=16.6cm,clip=,angle=-90]{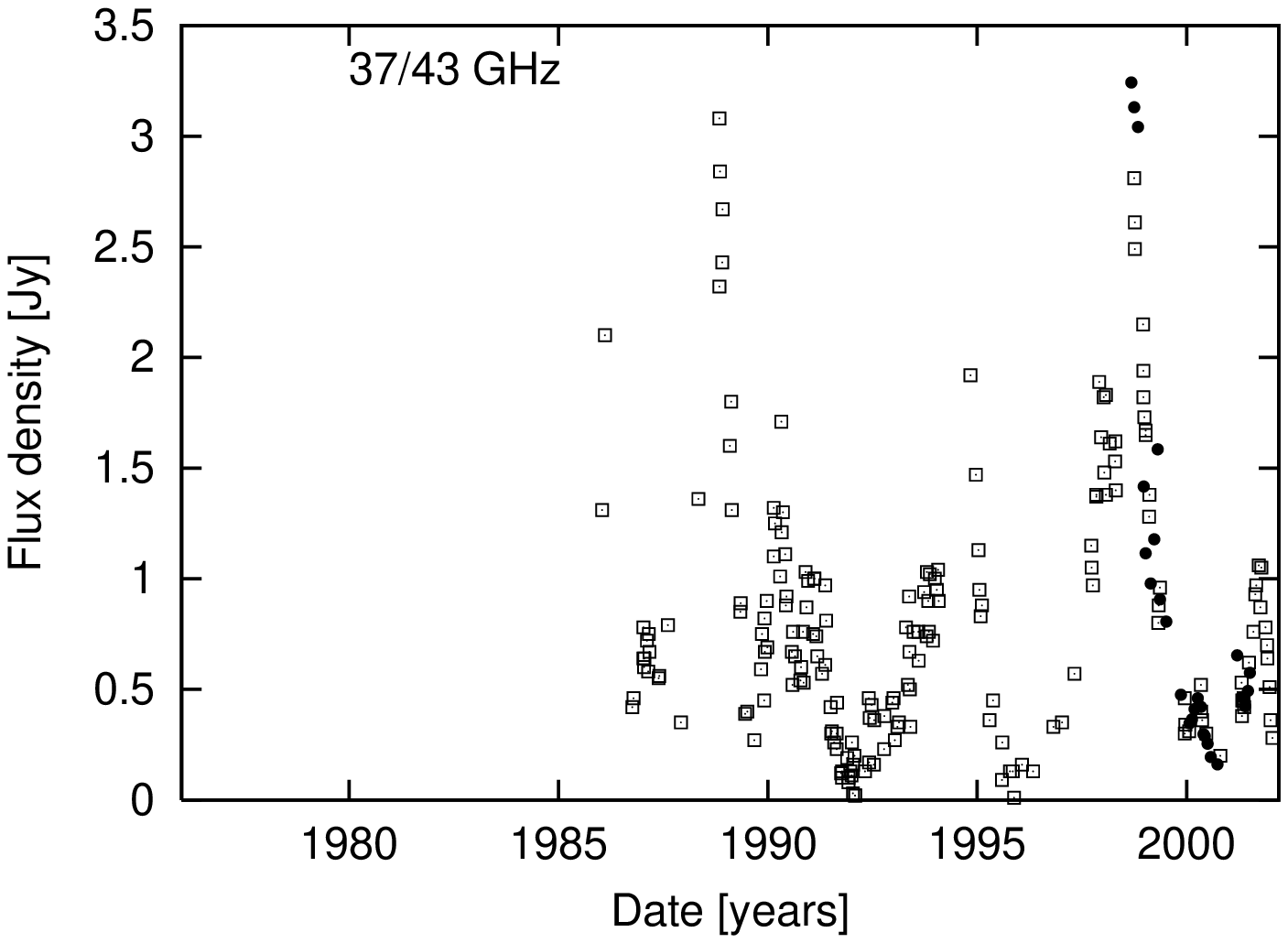}}
\caption{Radio light curves of III~Zw~2. The triangles at 8 and 15 GHz are from the Michigan monitoring program, the squares at 22 and 37 GHz are from the Mets\"ahovi monitoring program. The circles are our VLA observations at 8, 15, 22 and 43 GHz.}
\label{longterm}
\end{figure}

The core of III~Zw~2 shows extreme variability at radio wavelengths. Long
time radio light curves of this source spanning more than 20 years are shown
in Fig.~\ref{longterm} using data from Michigan, Mets\"ahovi, and the
VLA. One can see major flares with 30-fold increases in radio flux
density within two years. These major flares occur roughly every five
years with sub-flares on shorter timescales.

\begin{figure}
\resizebox{\hsize}{!}{\includegraphics[bbllx=9.2cm,bburx=18.2cm,bblly=1.9cm,bbury=16.6cm,clip=,angle=-90]{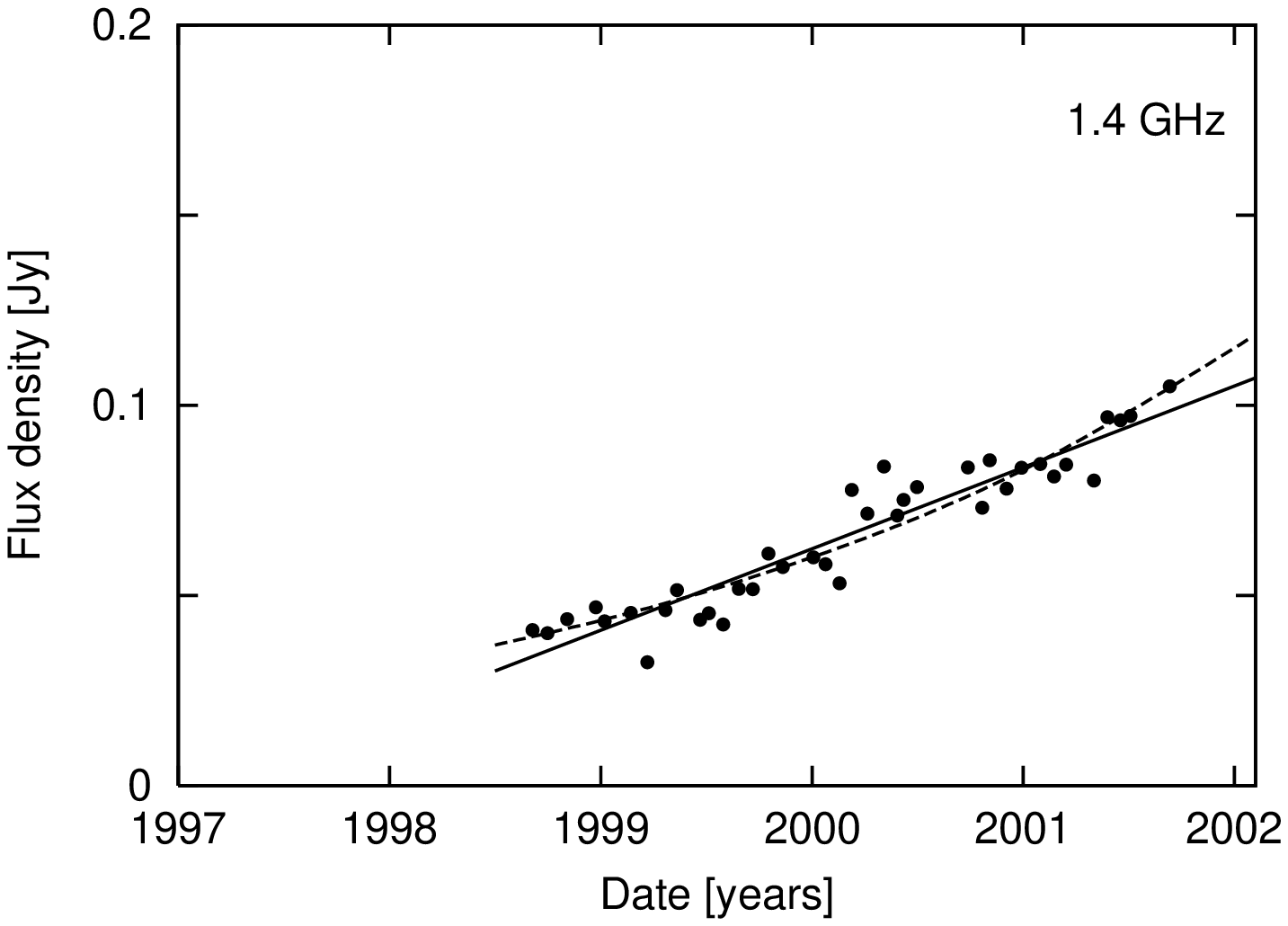}}
\resizebox{\hsize}{!}{\includegraphics[bbllx=9.2cm,bburx=18.2cm,bblly=1.9cm,bbury=16.6cm,clip=,angle=-90]{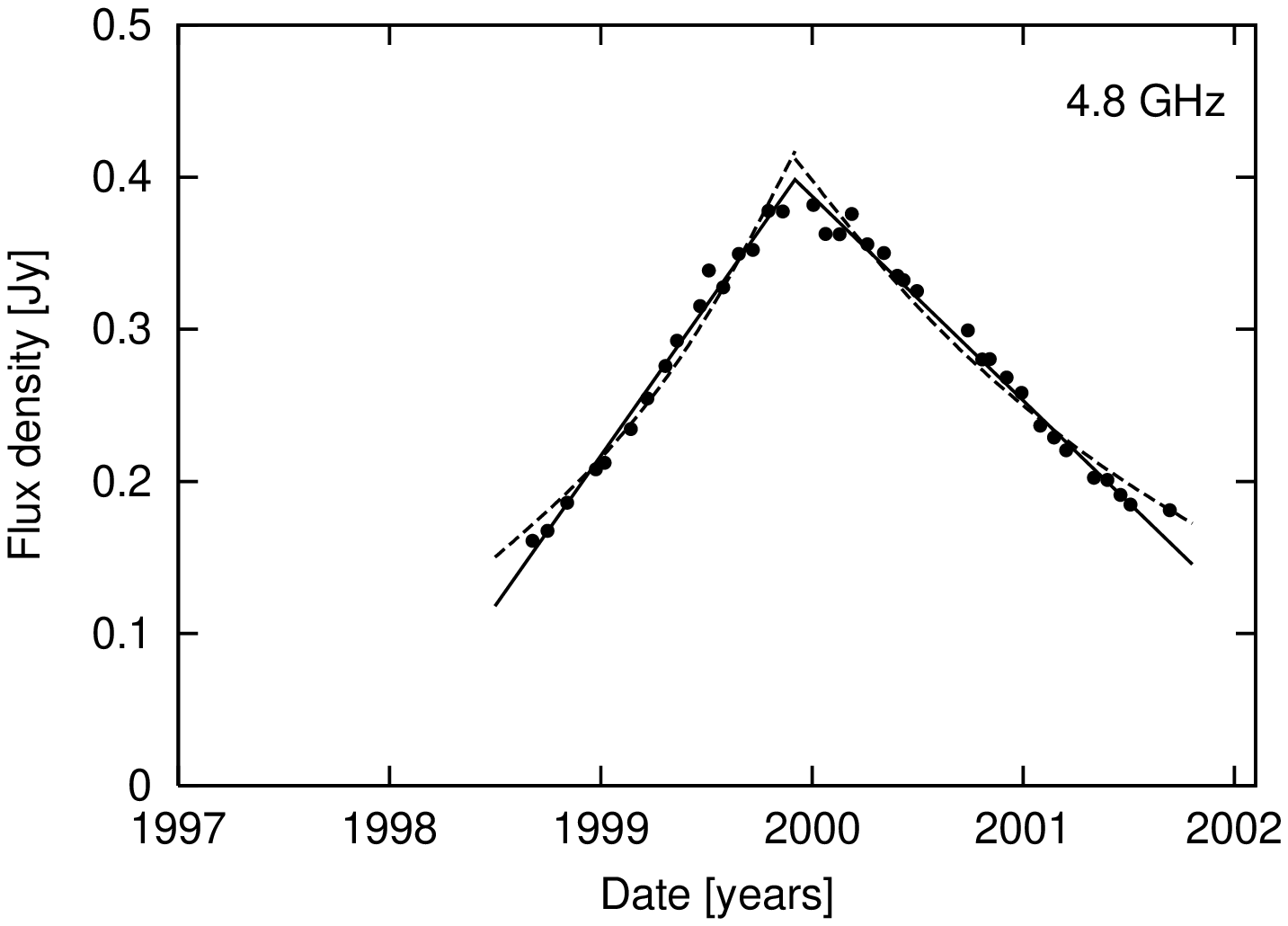}}
\resizebox{\hsize}{!}{\includegraphics[bbllx=9.2cm,bburx=19.6cm,bblly=1.9cm,bbury=16.6cm,clip=,angle=-90]{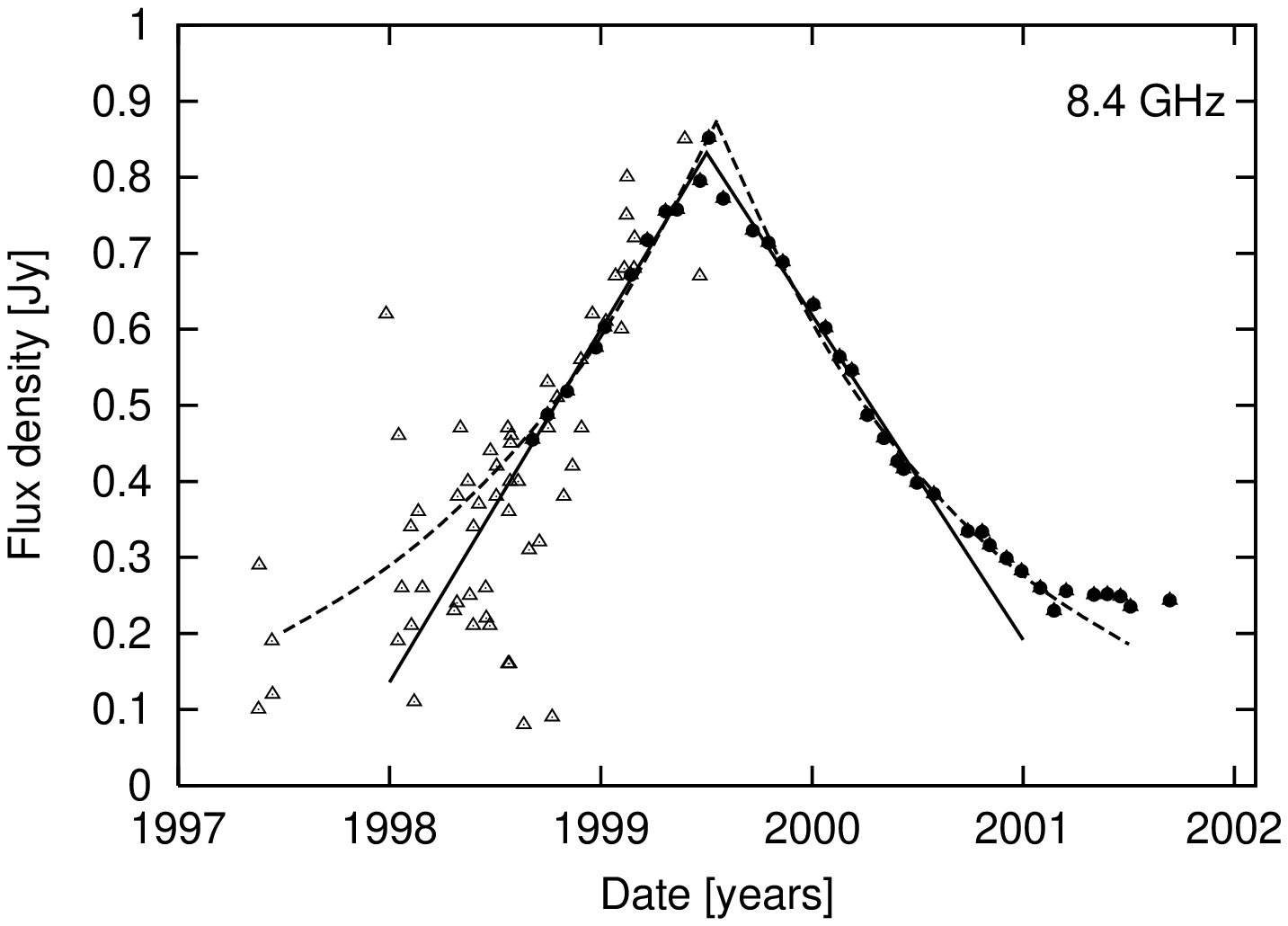}}
\caption{Radio light curves of the recent flare in III~Zw~2 at 1.4, 4.8 and 8.4 GHz. The circles are our VLA observations, and the triangles are from the Michigan monitoring program. The solid lines are our linear rise and decline fits. The dashed lines are the fitted exponential rise and decay.}
\label{flare1}
\end{figure}

\begin{figure}
\resizebox{\hsize}{!}{\includegraphics[bbllx=9.2cm,bburx=18.2cm,bblly=1.9cm,bbury=16.6cm,clip=,angle=-90]{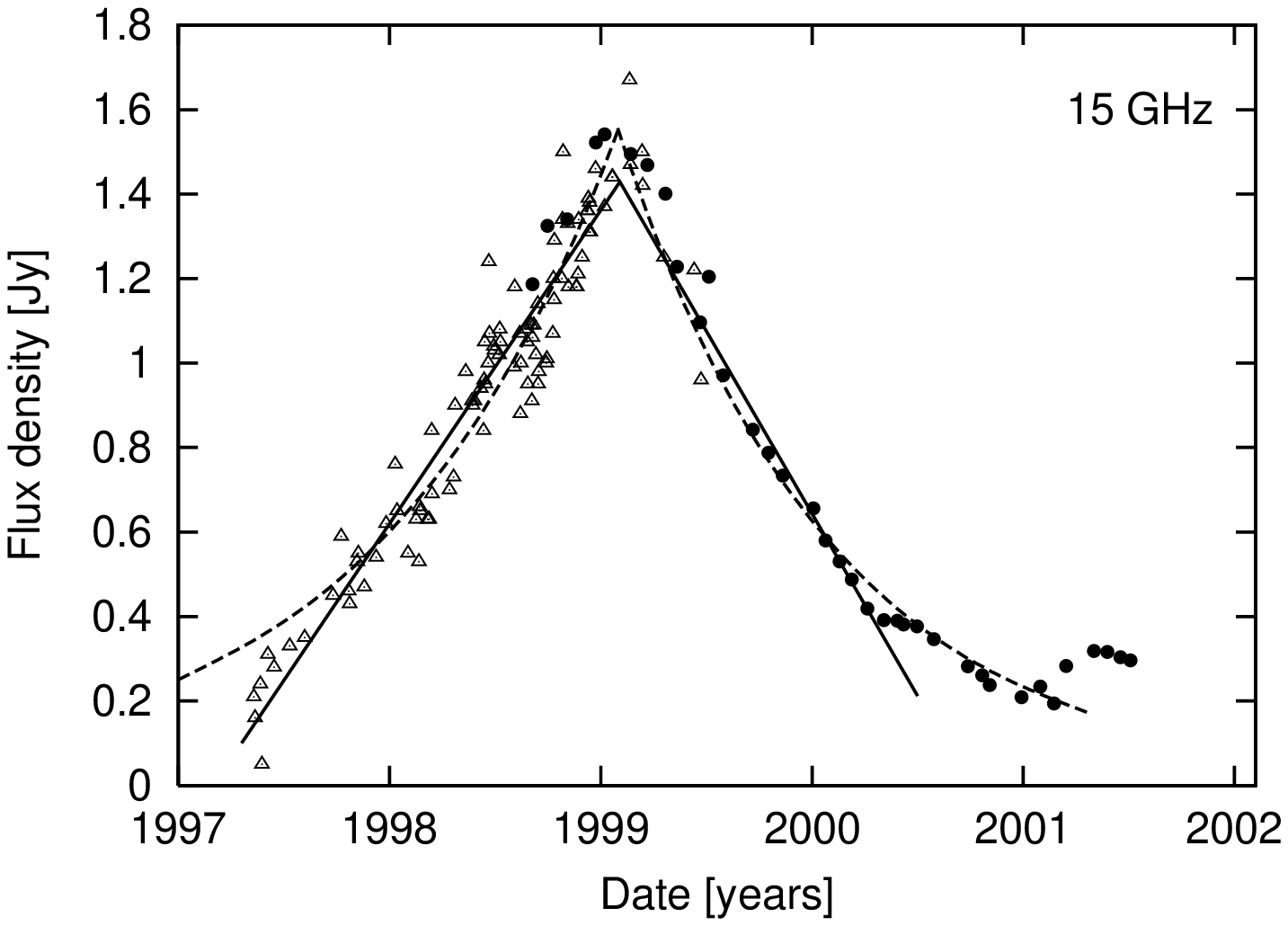}}
\resizebox{\hsize}{!}{\includegraphics[bbllx=9.2cm,bburx=18.2cm,bblly=1.9cm,bbury=16.6cm,clip=,angle=-90]{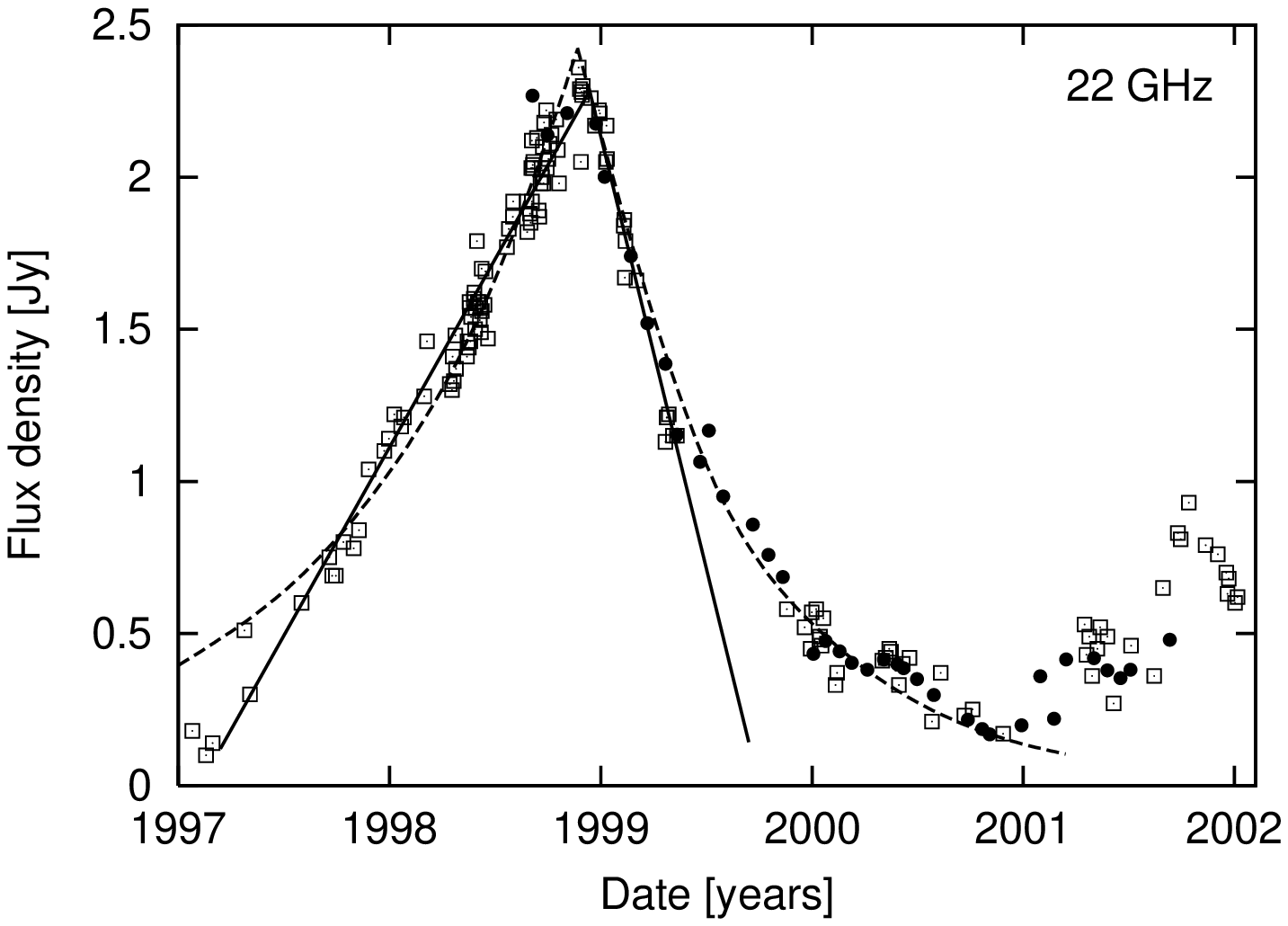}}
\resizebox{\hsize}{!}{\includegraphics[bbllx=9.2cm,bburx=19.6cm,bblly=1.9cm,bbury=16.6cm,clip=,angle=-90]{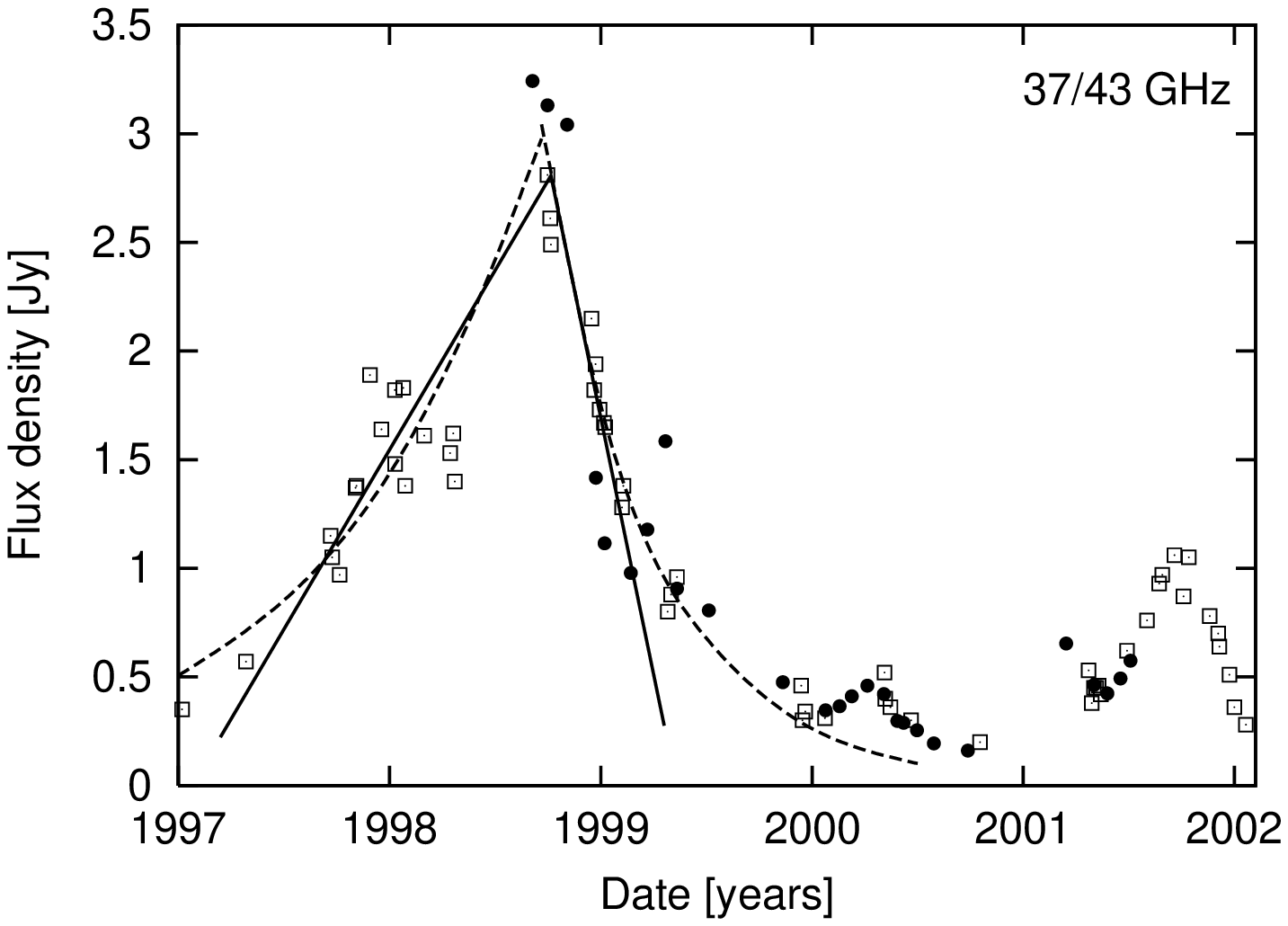}}
\caption{Radio light curves of the recent flare in III~Zw~2 at 15, 22 and 37/43 GHz. The circles are our VLA observations, and the triangles are from the Michigan monitoring program and the squares are from the Mets\"ahovi monitoring program. The solid lines are our linear rise and decline fits. The dashed lines are the fitted exponential rise and decay.}
\label{flare2}
\end{figure}

The outburst discussed in this paper started in 1996 and we monitored this
flare with the VLA. The good time sampling of one observation each month
allowed us to study this outburst in great detail. Lightcurves from the
most recent flare at six frequencies from 1.4 to 43 GHz are shown in
Fig.~\ref{flare1} and~\ref{flare2} together with our best model fits to
the data. Since the Michigan data at 8 GHz is rather noisy compared to
the other frequencies and the VLA data, we used
only our VLA monitoring data for the fits at this frequency. At 15, 22
and 37 GHz we used the VLA data as well as the Michigan and Mets\"ahovi
data. 

First we fitted a linear rise and decay to the flare. The rise is
consistent with a linear fit at all frequencies. The decay is also linear
at 4.8 GHz. At higher frequencies, the decay is linear only for a short
time period and deviates significant from a linear behavior at later times.
Thus we used only the linear part of the lightcurves for our fits.

The decay at higher frequencies can be fitted much better with an exponential
decay. Thus we fitted also an exponential rise and decay,
\begin{equation}
S(t)=S_0 e^{(t_0-t)/\tau_{r}},
\end{equation}

\begin{equation}
S(t)=S_0 e^{(t_0-t)/\tau_{d}},
\end{equation}
to the lightcurves. The fitting parameters are $S_0$, the maximum amplitude of
the flare and the flare rise and decay timescales $\tau_r$ and $\tau_d$. The
epochs of the flare maximum $t_0$ were taken from the observation with the
highest flux density. The increase of the lightcurves can also be fitted by an
exponential rise and the exponential decay fits the outburst until a new
smaller flare starts.

The slopes of the linear rise ($a_{r}$) and decay ($a_{d}$), the
exponential rise and decay timescales $\tau_r$ and $\tau_d$ and the
epoch of the flare maximum $t_0$ are listed in Table~\ref{lighttab} for
all six frequencies.

\begin{table}
      \caption[]{Fitting information: The slopes of the linear rise ($a_{r}$) and decay ($a_{d}$), the exponential rise and decay timescales $\tau_r$ and $\tau_d$ and the epoch of the flare maximum $t_0$.}
         \label{lighttab}
      \[
 \begin{tabular}{|c||cc|cc|c|}
           \hline
 Frequency  & $a_{r}$        & $a_{d}$        & $\tau_{r}$ & $\tau_{d}$  & $t_0$  \\
$[$GHz$]$&\multicolumn{2}{c|}{[Jy~yr$^{-1}$]}&\multicolumn{2}{c|}{[yr $^{-1}$]}&[yr]\\
            \hline
           1.4  & 0.021 &-&-3.088&-&- \\
           4.8  & 0.20 & -0.14&-1.39& 2.16 & 2000.1\\ 
           8.4  & 0.46 & -0.43&-1.40& 1.26&1999.5 \\ 
           15 & 0.74 & -0.86 &-1.14 &1.01 & 1999.1\\ 
           22  & 1.24& -2.85&-1.04 &0.73 &1998.9 \\ 
	   43  & 1.65&-4.73&-0.96 &0.53 & 1998.7\\ 
            \hline
         \end{tabular}
      \]
   \end{table}

\begin{figure}
\begin{center}
\includegraphics[width=0.35\textwidth,clip=,angle=-90]{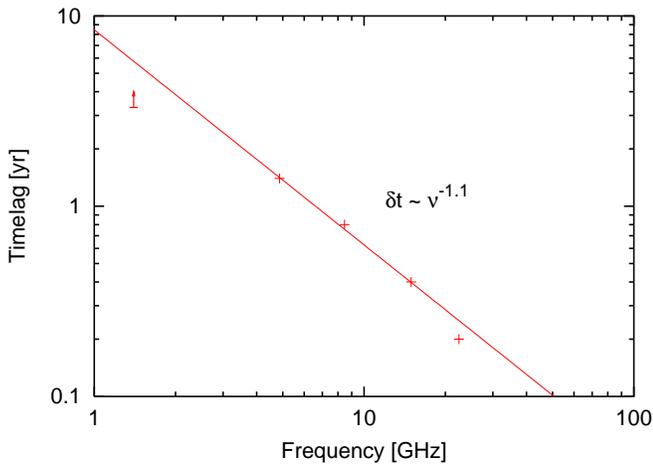}
\caption{Time-lag between the peak at 43 GHz and the peak at the other
  frequencies. The data point at 1.4 GHz is a lower limit.}
\label{time_peak}
\end{center}
\end{figure}

The lightcurve reaches its peak first at higher
frequencies. The time-lag between the peak at 43 GHz and the peak at
lower frequencies is shown in Fig.~\ref{time_peak} and is best fit
with a $\delta t \propto \nu^{-1.1}$ power law.
The rise and decay are faster at higher frequencies than at
lower frequencies. This behavior is typical for flares in AGN
(e.g.~\citeNP{TuerlerCourvoisierPaltani1999}). 

\begin{figure}
\begin{center}
\includegraphics[width=0.35\textwidth,clip=,angle=-90]{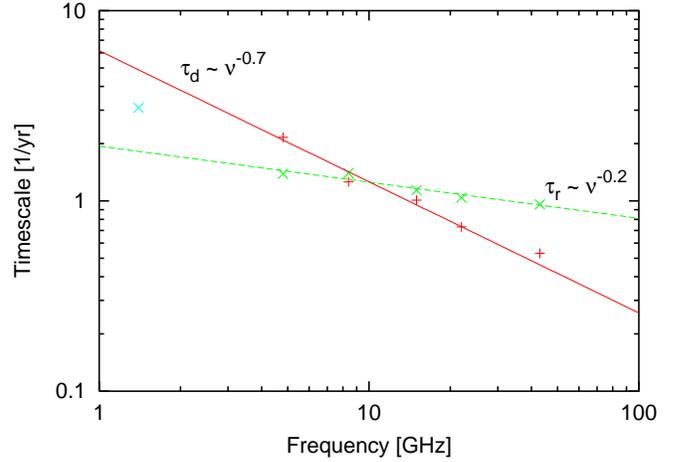}
\caption{Rise (x) and decay (+) timescales for all frequencies.}
\label{t_dr}
\end{center}
\end{figure}

At 4.8 GHz the rise is faster than the decay, while at 15, 22 and
43 GHz the rise is slower than the decay. This is the case for
the linear and the exponential fits to the flare. At 8.4 GHz the linear
rise is slightly slower than the linear decay while the exponential
rise slightly faster than the exponential decay is. The rise and decay
timescales are plotted as a function of frequency in
Fig.~\ref{t_dr}. The data can be fitted with power laws $\tau_r
\propto \nu^{-0.2}$ and $\tau_d \propto \nu^{-0.7}$. Only the rise
timescale at 1.4 GHz deviates from this power law. However, the
quiescence flux at 1.4 GHz is comparable to the flux density of the
outburst and will affect this data point.

The small dependancy of the rise timescale on frequency indicates that
optical depth effects are not very important during the rise. This is in
contrast to the decay, where optical depth effects are clearly important.

\citeN{ValtaojaLaehteenmaekiTeraesranta1999} modeled the radio lightcurves
of flares at 22 and 37 GHz in 85 extragalactic radio sources. They fitted an
exponential curve to the rise and the decay and found that in virtually all
flares a good model-fit could be obtained using a constant ratio between decay
and rise timescale of $\tau_{d}=1.3\tau_{r}$, i.e. the rise is faster than
the decay. These flares can be identified with the ejection of new VLBI
components in the jets.

The fact that the lightcurves of III~Zw~2 at 15, 22 and 43 GHz show the
opposite behavior with ratios of $\tau_d / \tau_r \approx 0.6 - 0.9$,
i.e. the decay is faster than the rise, indicates that
different physical processes are involved in this source. However, it
can not be excluded that the flare is composed of two closely spaced
flares.

\subsection{Spectral evolution}

During the outburst the spectrum of the source also showed variations. An
almost simultaneous broadband radio spectrum from 1.4 to 660 GHz during the
increase in flux density in May 1998 was presented in
~\citeN{FalckeBowerLobanov1999}. The spectrum was highly inverted at
centimeter wavelengths ($\alpha=+1.9\pm0.1$) with a turnover frequency
around 43 GHz. At frequencies above 43 GHz the spectrum became steep
with a spectral index of $\alpha=-0.75\pm0.15$, i.e. a textbook-like
synchrotron spectrum.

\begin{figure}
\resizebox{\hsize}{!}{\includegraphics[bbllx=9.2cm,bburx=18.2cm,bblly=1.9cm,bbury=16.6cm,clip=,angle=-90]{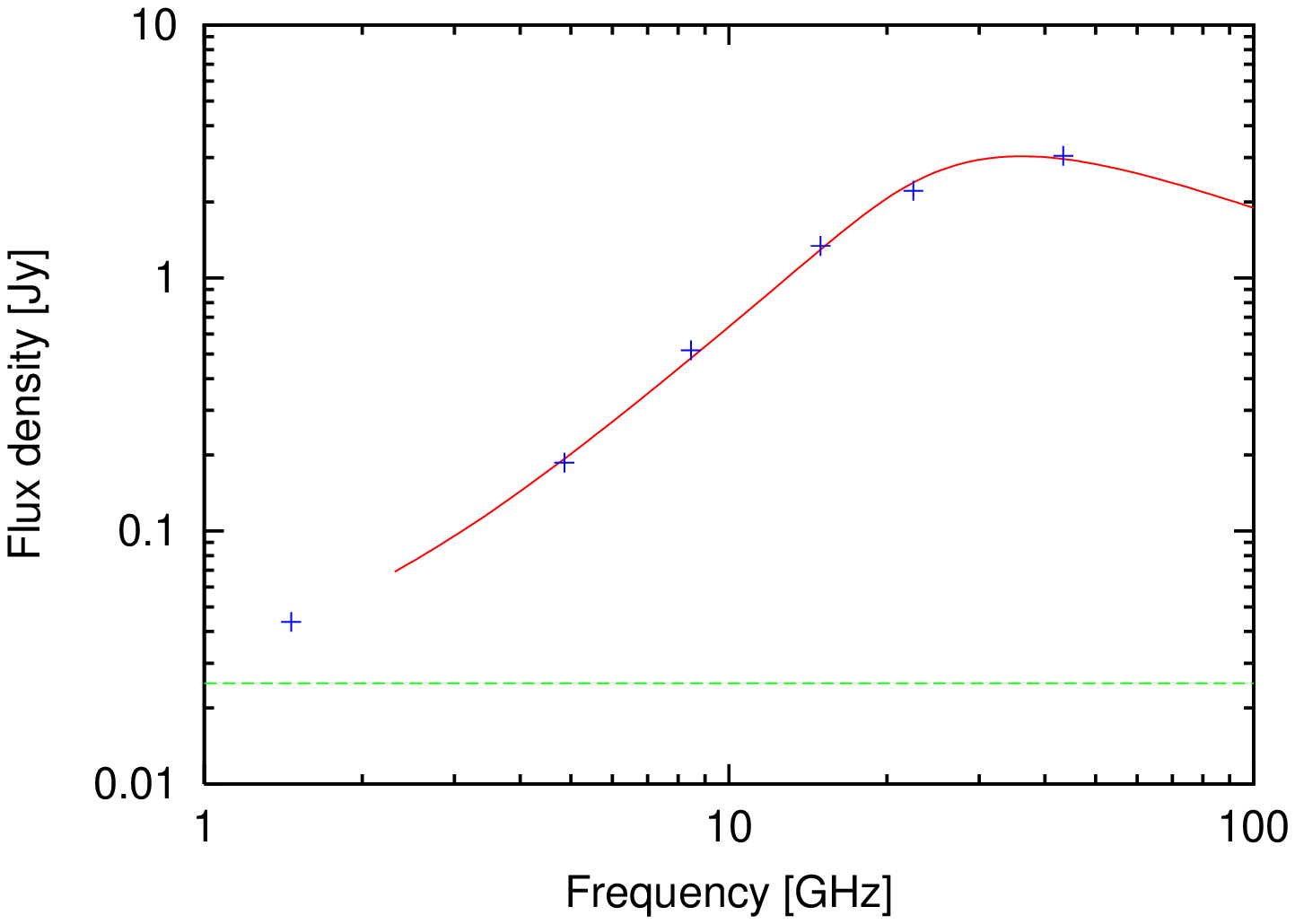}}
\resizebox{\hsize}{!}{\includegraphics[bbllx=9.2cm,bburx=18.2cm,bblly=1.9cm,bbury=16.6cm,clip=,angle=-90]{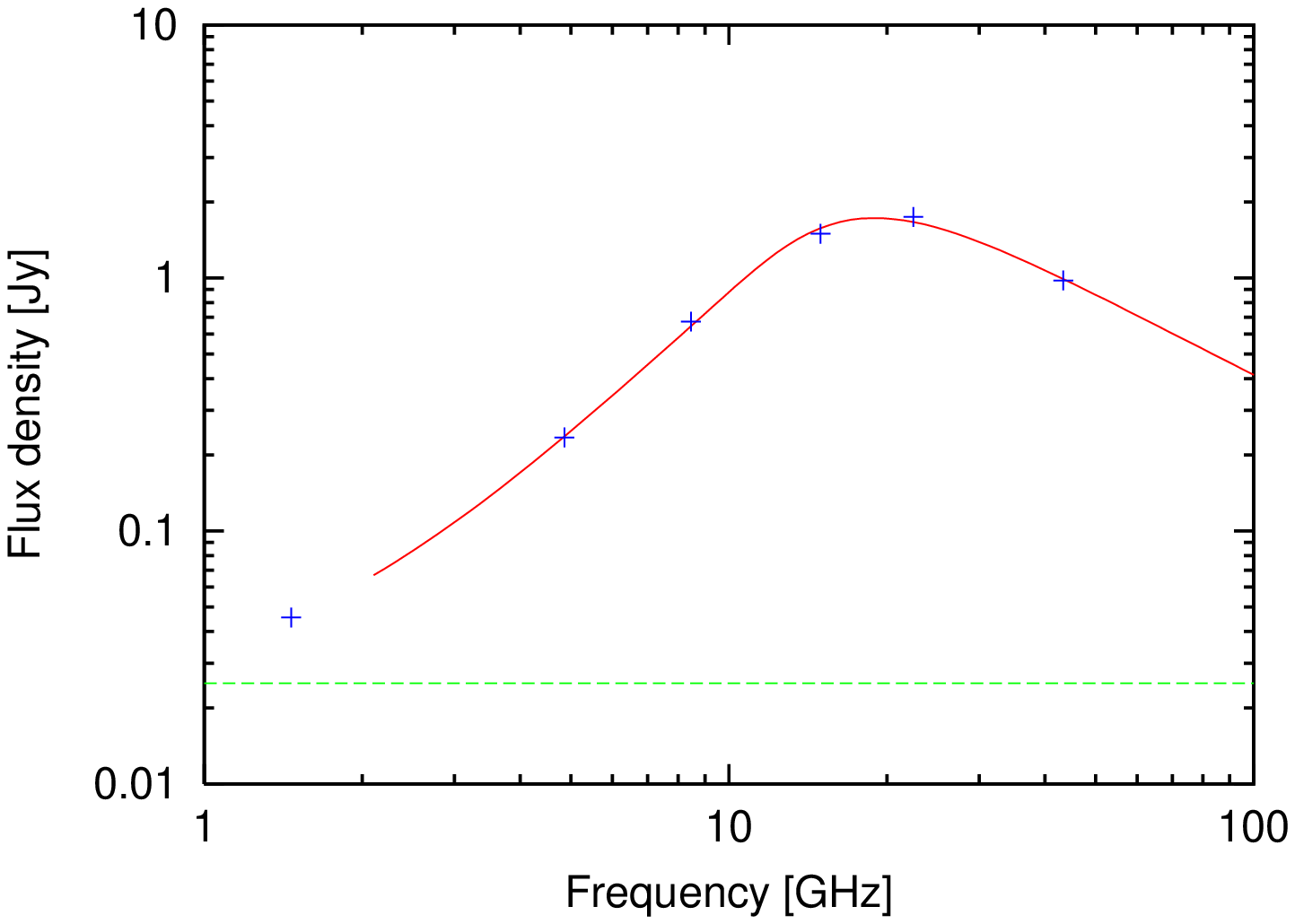}}
\resizebox{\hsize}{!}{\includegraphics[bbllx=9.2cm,bburx=18.2cm,bblly=1.9cm,bbury=16.6cm,clip=,angle=-90]{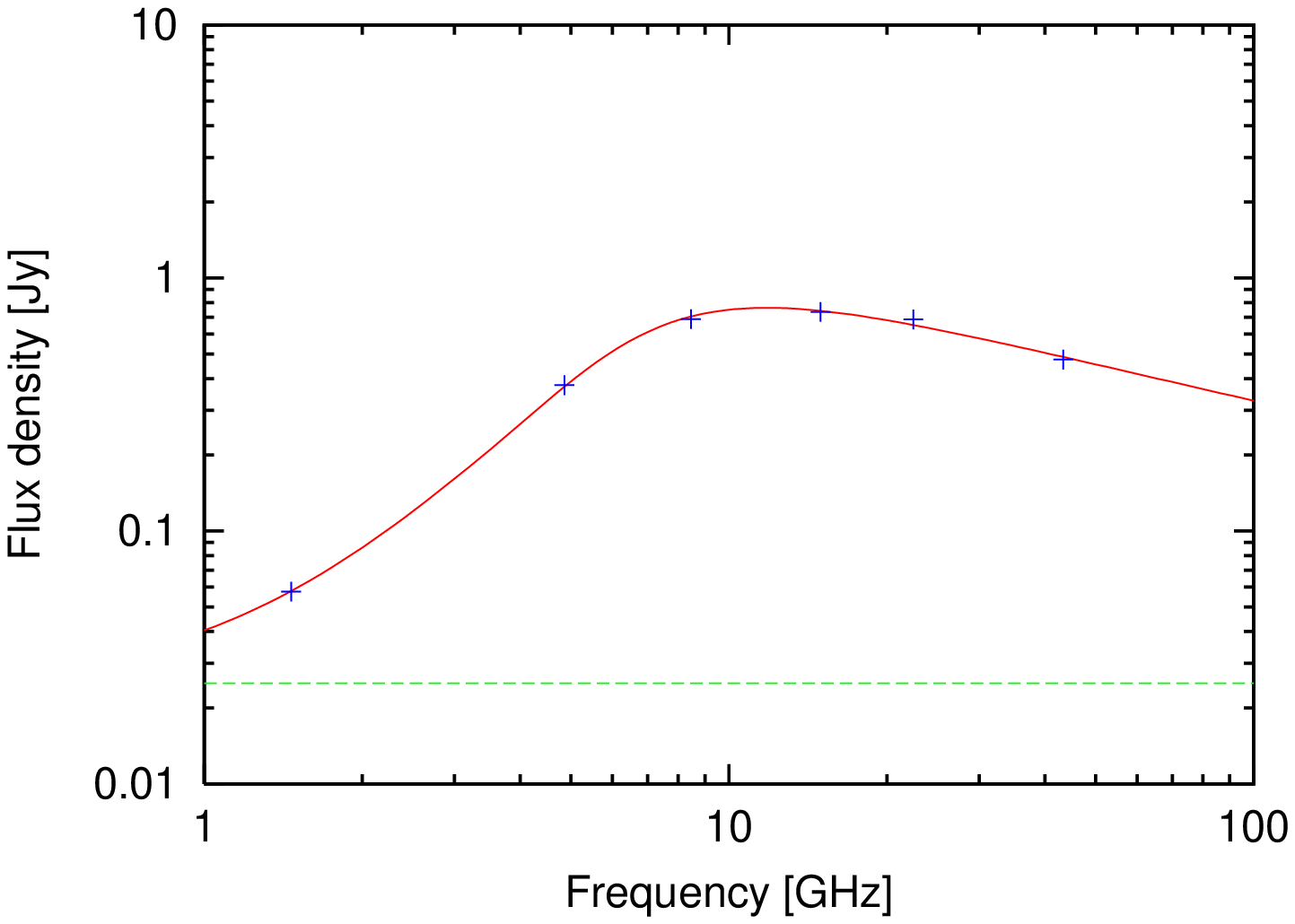}}
\resizebox{\hsize}{!}{\includegraphics[bbllx=9.2cm,bburx=19.6cm,bblly=1.9cm,bbury=16.6cm,clip=,angle=-90]{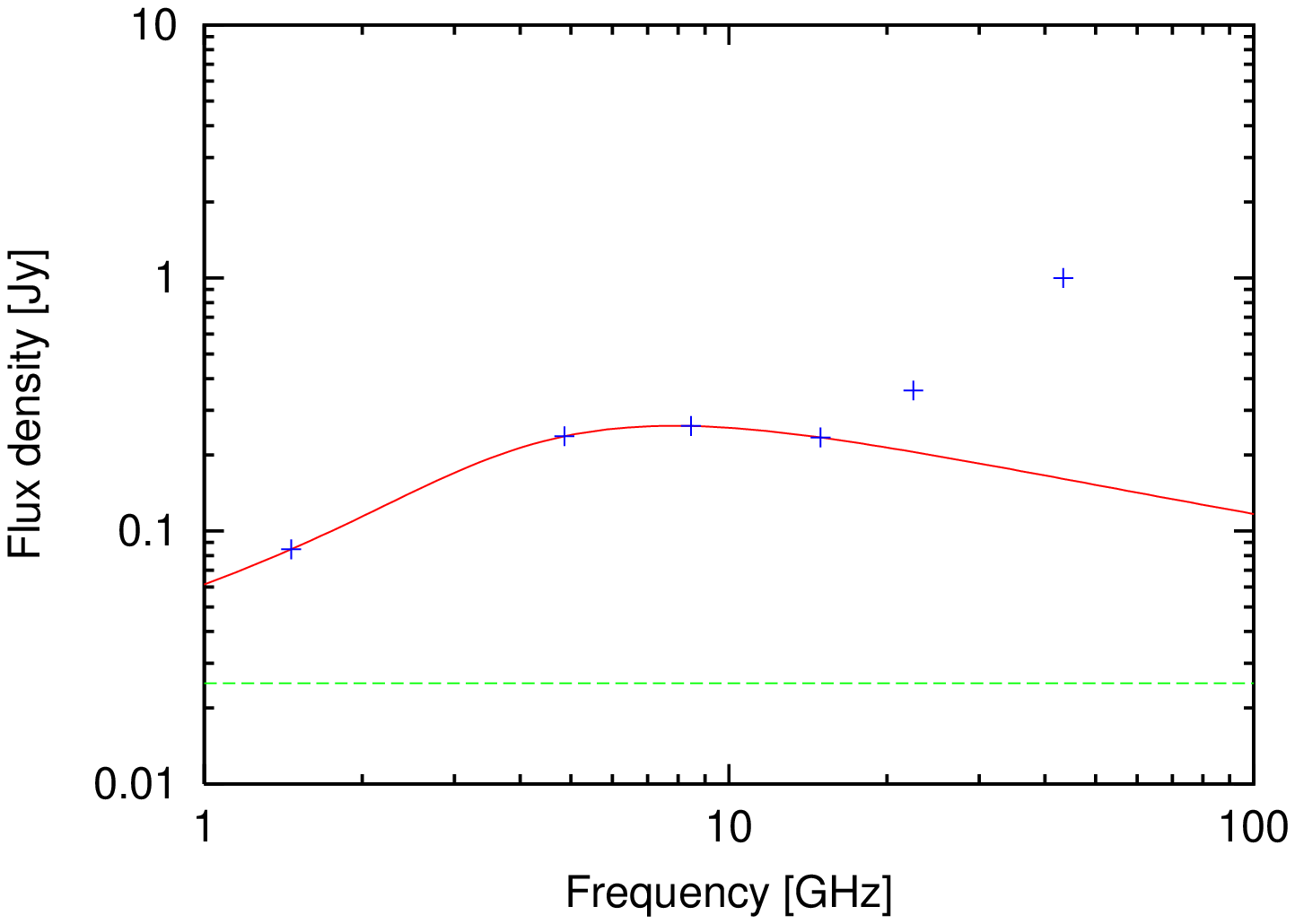}}
\caption{Spectra of III Zw 2 on 1998 November 04 (top), 1999 February
  22 (middle), 1999 November 12 (middle) and 2001 January 30 (bottom)
  together with the quiescent spectrum (horizontal line) and the best
  fit spectrum.}
\label{spec}
\end{figure}

Our VLA monitoring of the spectral evolution started in September 1998.
Four of the 41 epochs yielded no or bad data due to bad weather or hardware
failure. We fitted the remaining spectra with a broken power-law plus a flat
and constant quiescence spectrum $S_q$,
\begin{equation}
S(\nu)=S_{0} \left(\frac{\nu}{\nu_0}\right)^k\left(1-e^{-\left(\frac{\nu}{\nu_0}\right)^{l-k}}\right)+S_{q},
\end{equation}
where $k$ and $l$ are the spectral indices of the rising and declining
parts of the spectrum. $S_0$ and $\nu_0$ are fitting parameters and are
not exactly equal to the maximum flux density and the peak frequency of
the fitted spectrum.

We assume a flat spectrum for the quiescence flux which
is typical for quiescent cores in active galactic nuclei. Since we could fit
all epochs with a value of 25 mJy, we adopted this value for all epochs.

The approach with a broken power-law has the disadvantage, that there are
ambiguities in the parameters if the peak of the spectrum falls beyond the
frequencies covered in the observations.

This situation was the case in our first three observations in
September 1998, October 1998 and November 1998. The spectral
shape in these three epochs  was very similar to the May 1998 spectrum. Highly
inverted at centimeter wavelengths, a flattening towards higher frequencies,
and possibly a turnover around 43 GHz, but with higher flux densities (see
Fig.~\ref{spec}, top).
Thus we assumed the spectral index of the declining part of the spectrum to be
$l=-0.75$, the value of the May 1998 observation. This is a reasonable
assumption since the overall spectral shape did not change
significantly between May 1998 and September 1998.

\begin{figure}
\resizebox{\hsize}{!}{\includegraphics[bbllx=9.2cm,bburx=19.6cm,bblly=1.9cm,bbury=16.9cm,clip=,angle=-90]{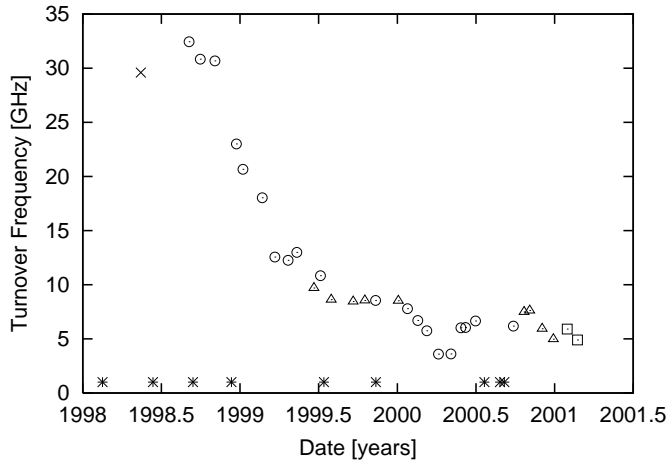}}
\caption{Evolution of the "turnover frequency" $\nu_0$. The circles
  mark epochs with 6 observed frequencies (1.4-43 GHz) and the
  triangles epochs with 5 observed frequencies (1.4-22 GHz). In the
  last two epochs only 4 frequencies (1.4-15 GHz) were used because of
  a new outburst at high frequencies. The cross is the turnover
  frequency of the spectrum in May 1998, where the 1-43 GHz data is
  taken from \protect\citeN{FalckeBowerLobanov1999}. The asterisks mark
  the epochs of our VLBA observations.}
\label{turnover}
\end{figure}

\begin{figure}
\resizebox{\hsize}{!}{\includegraphics[bbllx=9.2cm,bburx=19.6cm,bblly=1.9cm,bbury=16.9cm,clip=,angle=-90]{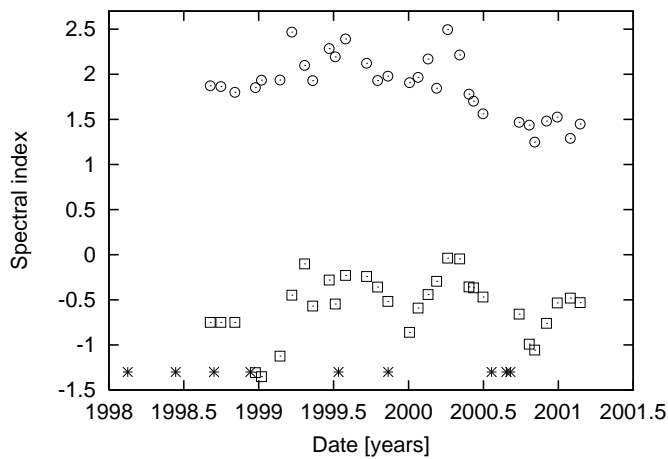}}
\caption{Evolution of the spectral indices k (circles) and l (squares). The asterisks mark the epochs of our VLBA observations.}
\label{kandl}
\end{figure}

The temporal evolution of the fitting parameters $\nu_0$, $k$ and $l$ can be
seen in Fig.~\ref{turnover} and Fig~\ref{kandl}. In some epochs we covered
only 5 frequencies from 1.4 to 22 GHz. The absence of
the 43 GHz flux density in these epochs could bias the results of the
spectral fitting. Thus we marked the epochs with only 5 frequencies in
Fig.~\ref{turnover} with triangles while the epochs with 6 frequencies are
indicated by circles. One can see that the fits of both subsets are in good
agreement.

After November 1998 the spectrum underwent a dramatic change (see also
Fig.~\ref{spec}). The turnover frequency $\nu_0$, that stayed roughly 
constant at around 30 GHz from May 1998 until November 1998 dropped to
23 GHz in December 1998. In the following months, the turnover drops
further until it reached 10 GHz in June 1999. During the next year the
turnover frequency showed only smaller and slower variations and stayed
roughly constant at $\sim 7$~GHz. The drop in turnover frequency to 4
GHz in the first months of 2000 can be explained by the onset of a new
minor flare at high frequencies (see 43 GHz lightcurve in
Fig.~\ref{flare2}). The new flare caused a flattening of the optical
thin part of the spectrum and a shift of the turnover to lower
frequencies. The flattening can also be seen in Fig.~\ref{kandl} where
the spectral index $l$ changes from $\sim -0.9$ to $\sim -0.1$ during
that time. 

The spectral index in the optical thick part of the spectrum $k$ stays at
$\sim 2$ and slowly flattens towards later times.

A new strong outburst started at high frequencies in January 2001 (see 22 and
43 GHz lightcurves in Fig.~\ref{flare2}) and one would have to model two
independent broken power-laws to the spectrum. Since one broken power-law
is characterized by four parameters $S_0$, $\nu_0$, $k$ and $l$,
our six data points in each spectrum are not sufficient to model two
independent components with four parameters each. In the first two epochs of
the new flare, only the 22 and 43 GHz data were affected and we fitted the
broken power-law to the remaining four frequencies.

The fast change in peak frequency implies also a strong morphological change,
i.e. a rapid expansion. This prediction was tested by VLBI observations which
are described in the next section.

\subsection{Structural evolution}
\label{vlbi}    
\subsubsection{43 GHz Results}

\begin{figure}
\resizebox{\hsize}{!}{\includegraphics[bbllx=3.3cm,bburx=16.2cm,bblly=9.8cm,bbury=27.8cm,clip=,angle=0]{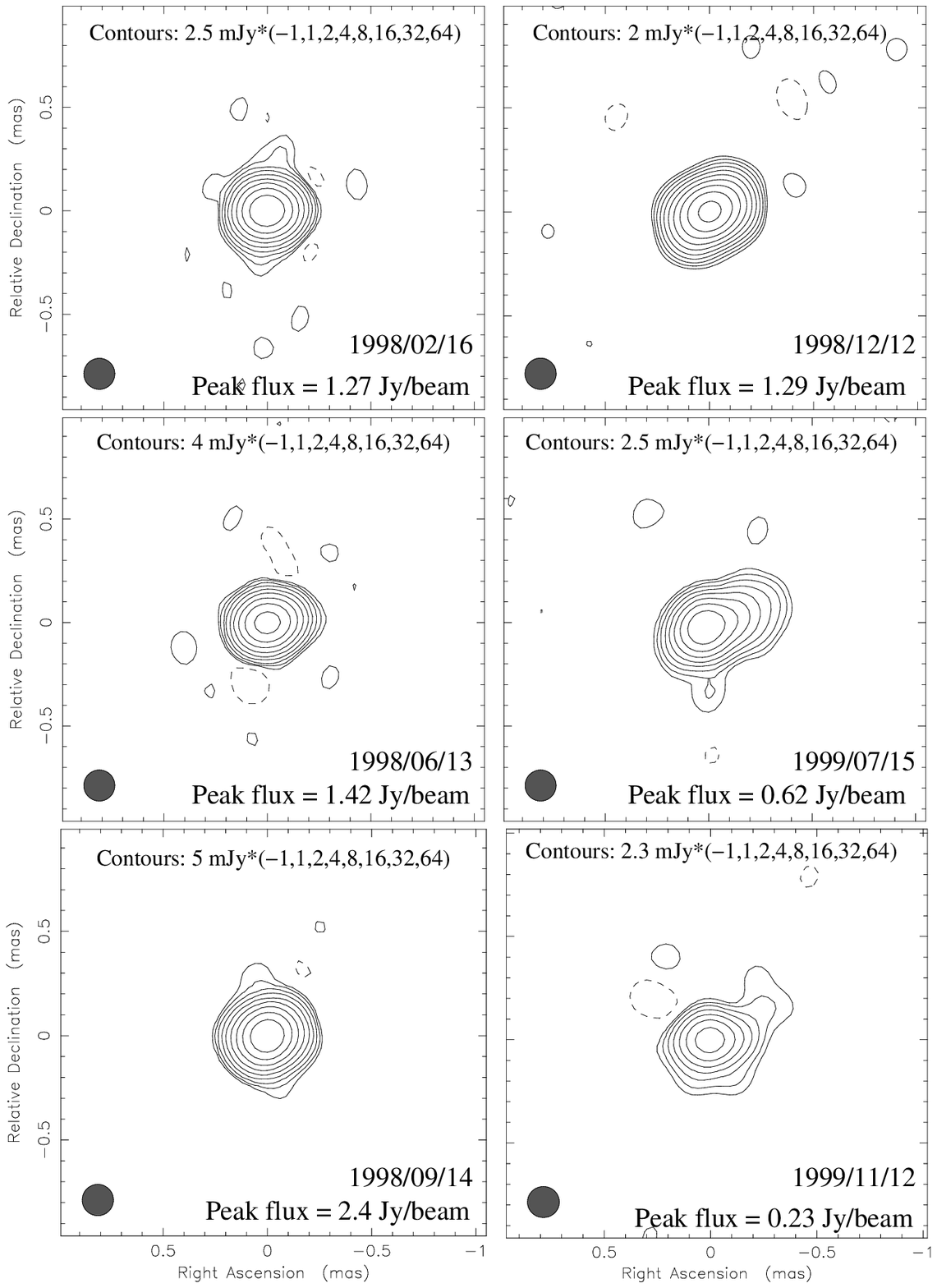}}
\caption{All six VLBA maps of III~Zw~2 at 43 GHz convolved with a superresolved circular beam of 150 $\mu$as.}
\label{43maps}
\end{figure}

The first three VLBA observations were made during the first phase of the
flare, marked by the increase in flux density and a roughly constant spectral
peak above 30 GHz. The constancy of the peak frequency indicates no structural
change, since the turnover is caused by synchrotron self-absorption
(\citeNP{FalckeBowerLobanov1999}). The source is slightly resolved and
the long baselines show non-zero closure phases, indicating an
asymmetric structure. Two point-like components were fitted to the
uv-data to represent the extent of the source. The source shows no
structural change (see Fig.~\ref{43maps}) and the separation of the two
components during this phase of the flare stayed constant at $\sim
76~\mu as$, corresponding to $\sim 0.11$~pc. The excellent agreement
within 2~$\mu$as between the first three epochs shows the high quality
of the data and the accuracy of the relative astrometry.

After November 1998, the VLA monitoring shows a dramatic change in the
spectrum. The peak frequency dropped quickly to 10 GHz within a few
month (Fig.~\ref{turnover}). In the framework of a simple equipartition
jet model with a $R\propto \nu_{\mathrm{ssa}}^{-1}$ dependence (e.g.,
\citeNP{BlandfordKonigl1979}; \citeNP{FalckeBiermann1995}) one would
expect a rapid expansion. With a source size of 0.11 pc and a turnover
frequency of 33 GHz in the first phase with no expansion one expects a
source size of 0.36 pc for a self-absorption frequency of 10 GHz
according to the spectral evolution.

   \begin{table}
      \caption[]{Flux densities S, separation D and position angle P.A. of the
        two outermost point-like components of our model-fits to the 43 GHz uv-data.}
         \label{7mm_tab}
      \[
  \begin{tabular}{p{0.18\linewidth}|p{0.12\linewidth}p{0.12\linewidth}p{0.12\linewidth}cp{0.10\linewidth}c}
           \hline
 Date & $S_1$ [Jy] &  $S_2$ [Jy] & $S_3$ [Jy]& d [mas] & P.A.\\
            \hline
           1998/02/16 & 0.93 & 0.58 & -- & 0.075 &  $-84^{\circ}$\\
           1998/06/13 & 1.03 & 0.65 & -- & 0.077 &  $-78^{\circ}$\\
           1998/09/14 & 1.60 & 1.27 & -- & 0.077 &  $-72^{\circ}$\\
           1998/12/12 & 0.86 & 0.86 & -- & 0.106 &  $-63^{\circ}$\\
           1999/07/15 & 0.56 & 0.26 & 0.08& 0.245 &  $-71^{\circ}$\\
	   1999/11/12 & 0.08 & 0.23 & 0.05 & 0.246 &  $-73^{\circ}$\\
            \hline
         \end{tabular}
      \]
   \end{table}

Indeed, the fourth VLBA epoch, observed only one month after the start
of the spectral evolution, shows first signs of an expansion. The fifth
epoch shows a dramatic structural change (see Fig.~\ref{43maps}) and a
model of three point-like components is required to describe the data
now. The separation between the outer components is now $\sim 245~\mu
as$ corresponding to $\sim 0.37$~pc. This is in good agreement with the
expected value of 0.36 pc from the equipartition jet model. The
structure in the sixth epoch is very similar to the fifth epoch but
with lower flux density. This is again expected, since the turnover
frequency stayed around 10 GHz.

The separation of the outer components for all six epochs is plotted in
Fig.~\ref{expand1} (upper panel). For the first three epochs we measure
an upper limit for the expansion speed of 0.04 $c$. The rapid expansion
between the fourth and fifth epoch shows an apparent speed of 1.25
$c$. Between the last two epoch we detected again no expansion with an
upper limit of 0.04 $c$.

Unfortunately the source was too weak at 43 GHz to be detected at the
last three epochs.

\subsubsection{15 GHz Results}

\begin{figure}
\resizebox{\hsize}{!}{\includegraphics[bbllx=9.2cm,bburx=18.2cm,bblly=1.9cm,bbury=16.6cm,clip=,angle=-90]{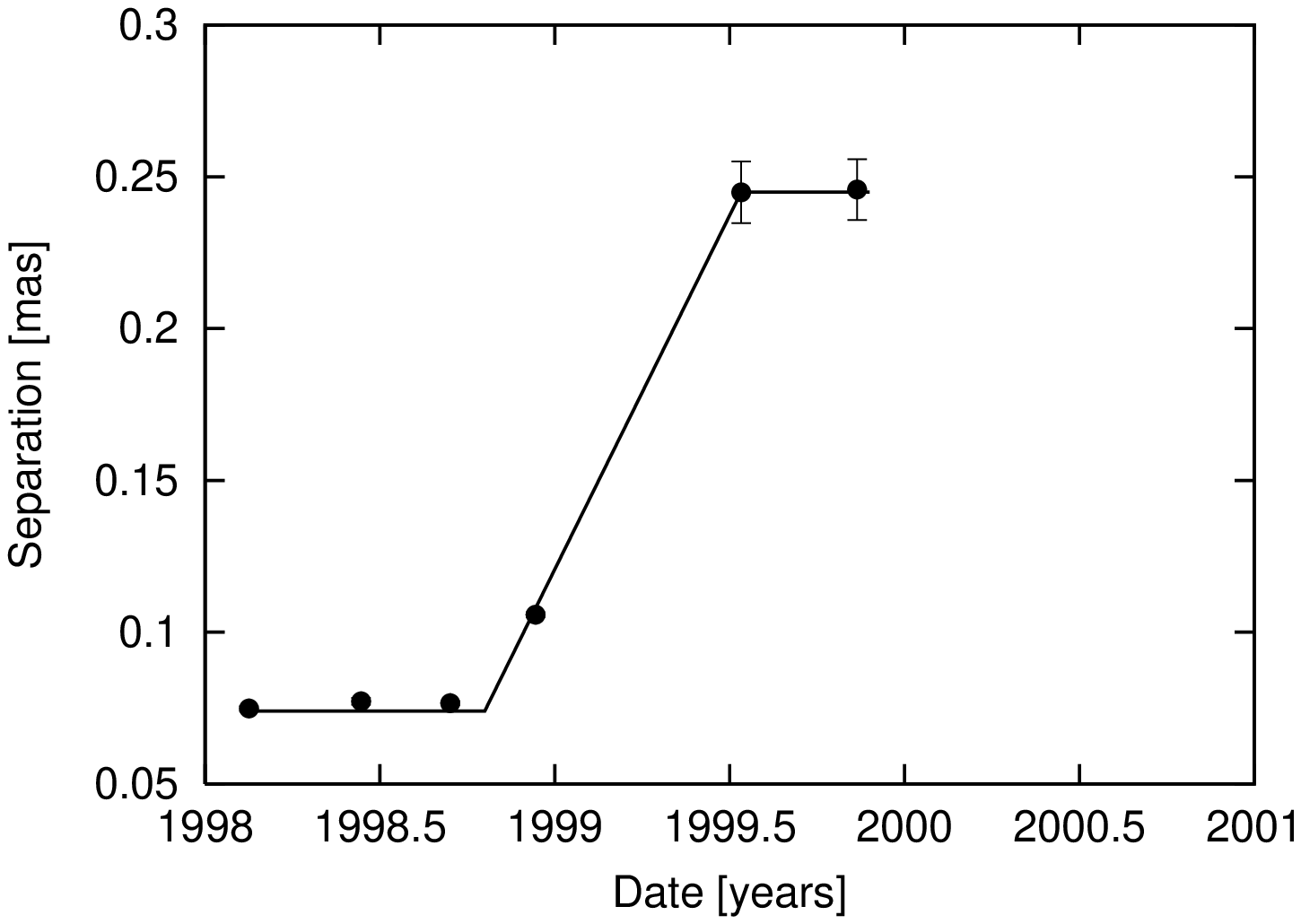}}
\resizebox{\hsize}{!}{\includegraphics[bbllx=9.2cm,bburx=19.6cm,bblly=1.9cm,bbury=16.6cm,clip=,angle=-90]{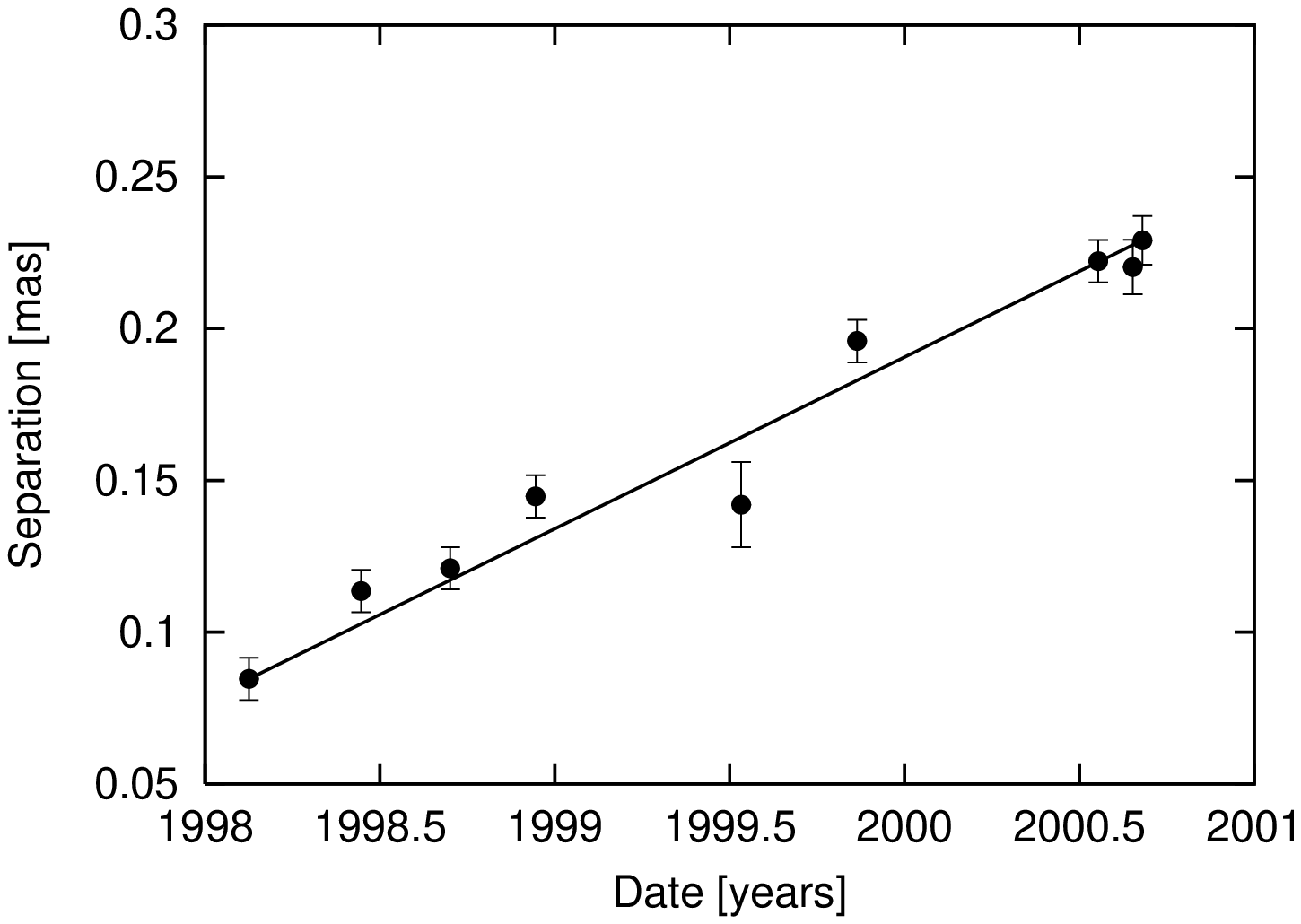}}
\caption{Component separation at 43 (upper) and 15 (lower) GHz. The apparent expansion speed is 0.6 c at 15 GHz and 1.25 c at 43 GHz.}
\label{expand1}
\end{figure}

At 15 GHz the picture looks completely different. The source is very
compact but slightly resolved in all epochs except the first and we
fitted two point-like components to the uv-data. The flux densities,
separations and position angles of the two components are listed in
Table~\ref{2cm_tab}. The component separation of all epochs is also
plotted in Fig.~\ref{expand1}. One can see a constant expansion with an
apparent expansion speed of $\sim0.6~c$. Simple extrapolation backwards
suggests that the expansion has started in May 1996. This is consistent
with the onset of the new flare in the 37 GHz lightcurve in
Fig.~\ref{longterm}. The flare started between two 37 GHz observations
in May and October 1996.

One should note that the component separation of the fifth epoch shows a
deviation from a constant expansion. If one splits up the separation into
its north-south and east-west components the scatter is larger in the
north-south direction. This is expected, since the beam of the VLBA is
elongated in the north-south direction. This relatively large scatter
in north-south direction also explains the scatter in the positions
angles in Table.~\ref{2cm_tab}. 



   \begin{table}
      \caption[]{Flux densities S, separation D and position angle P.A. of the
        two point-like components of our model-fits to the 15 GHz uv-data.}
         \label{2cm_tab}
      \[
  \begin{tabular}{p{0.18\linewidth}|cp{0.12\linewidth}cp{0.12\linewidth}cp{0.10\linewidth}cp{0.10\linewidth}c}
           \hline
 Date & $S_1$ [Jy] &  $S_2$ [Jy] &d [mas] & P.A.\\
            \hline
           1998/02/16 & 0.32 & 0.40 & 0.085 &  $-54^{\circ}$\\
           1998/06/13 & 0.45 & 0.48 & 0.114 &  $-79^{\circ}$\\
           1998/09/14 & 0.67 & 0.50 & 0.121 &  $-70^{\circ}$\\
           1998/12/12 & 0.79 & 0.69 & 0.145 &  $-59^{\circ}$\\
           1999/07/15 & 0.50 & 0.46 & 0.142 &  $-87^{\circ}$\\
	   1999/11/12 & 0.43 & 0.25 & 0.195 &  $-68^{\circ}$\\
	   2000/07/22 & 0.22 & 0.10 & 0.222 &  $-61^{\circ}$\\
	   2000/08/27 & 0.19 & 0.08 & 0.220 &  $-68^{\circ}$\\
	   2000/09/06 & 0.18 & 0.07 & 0.229 &  $-63^{\circ}$\\
            \hline
         \end{tabular}
      \]
   \end{table}

\section{Discussion}
The stop-and-go behavior and the apparent contradiction between the 43
GHz and 15 GHz data can be explained by a jet interacting with the
interstellar medium in combination with optical depth effects in an
'inflating-balloon model'.

In this model, the initial phase of the flare can be explained by a
relativistic jet interacting with the interstellar medium or a torus
that creates a shock and gets frustrated. A relativistic shock was
proposed by \citeN{FalckeBowerLobanov1999} due to synchrotron cooling
times of 14-50 days which are much shorter than the duration of the
outburst. The ultra-compact hotspots are pumped up, powered by the jet
and responsible for the increase in flux density. The post-shock
material expands with the maximum sound speed of a magnetized
relativistic plasma of $c_s\approx 0.6$~c.

Since the source is optically thick at 15 GHz, one observes the outside
of the source, i.e. the post-shock material expanding with sound
speed. At 43 GHz, the source is optically thin and one can look inside
the source and see the stationary hotspots.

The rapid expansion at 43 GHz thereafter has marked the phase where the
jet breaks free and starts to propagate relativistically into a
lower-density medium. Then the expansion stops again when the jet hits
another cloud.

The fact that spectral and structural evolution during the outburst are
closely linked demonstrates that we are dealing with a real physical
expansion and not only a phase velocity. The observations described
here produced a large amount of data that are all consistent with a
simple synchrotron self-absorbed jet model.

For the question of the nature of the radio-loud/radio-quiet dichotomy
this means that radio-weak and radio-loud quasars can indeed have
central engines that are in many respects very similar. Their optical
properties are almost indistinguishable and both types of quasars can
produce relativistic jets in their nuclei. The finding of superluminal
motion supports the hypothesis of \citeN{MillerRawlingsSaunders1993} and
\citeN{FalckePatnaikSherwood1996} that RIQs are relativistically
boosted intrinsically radio-weak AGN. Recently, a further relativistic
jet in a radio-quiet quasar was found by~\citeN{BlundellBeasleyBicknell2003}.

However, the nature of the medium interacting with the jet remains
unclear. The  outbursts could be explained by a
precessing jet that hits a molecular torus roughly every five years. So
far, no direct evidence for molecular gas in the nucleus of III~Zw~2
was found.

Some Seyfert galaxies have shown H$_2$O maser emission associated with
the nuclear jet. In these sources, the maser emission is the result
from an interaction of the jet with a molecular cloud. One example is
the Seyfert II galaxy Mrk\,348 (see
\citeNP{PeckHenkelUlvestad2003}). In this source, the ejection of a new
VLBI component has lead to a flare of the radio source similar to the
outburst in III~Zw~2. The outburst started with a peak frequency of
$\approx$ 22 GHz which gradually decreased over 20 months. During this
outburst, H$_2$O maser emission was found (\citeNP{FalckeHenkelPeck2000}). Searches for similar water
maser emission in III~Zw~2 with the Effelsberg 100-m telescope yielded
no detection (Henkel, private communication).

\begin{figure}
\resizebox{\hsize}{!}{\includegraphics[bbllx=9.1cm,bburx=20.2cm,bblly=1.7cm,bbury=16.5cm,clip=,angle=-90]{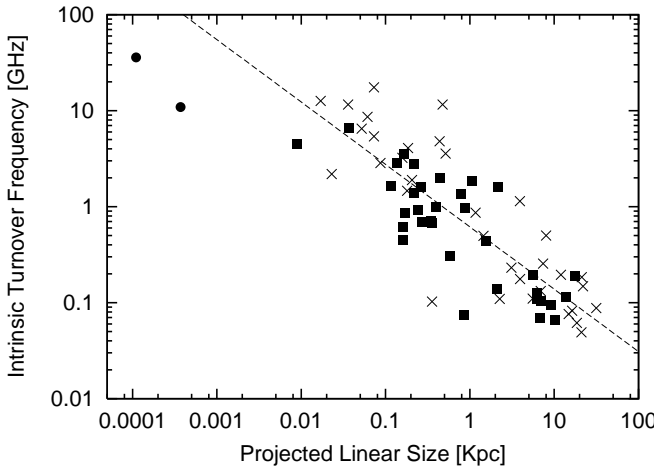}}
\caption{Intrinsic turnover frequency vs. linear size for GPS and CSS sources. The quasars are represented by crosses, and the galaxies by solid squares. Adapted from~\protect\citeN{ODeaBaum1997}. The two circles mark the values for III~Zw~2 before and after the expansion at 43 GHz.}
\label{nu_s}
\end{figure}

In the currently favoured youth model for Compact Steep Spectrum (CSS) and 
GHz Peaked Spectrum (GPS) sources, the linear size of a source is related 
to the age of the source. The correlation between the turnover frequency and 
the projected linear size (e.g.,~O'Dea \& Baum 1997) suggests that the 
turnover frequency decreases while the source ages and expands. Therefore
the sources with the highest turnover frequencies represent the youngest 
objects. In Fig.~\ref{nu_s} we plot linear size vs.  turnover frequency
for GPS and CSS sources. We include III~Zw~2  before ($\nu_{ssa}\approx
33~$GHz; size $\approx$~0.11~pc) and after  ($\nu_{ssa}\approx 10~$GHz;
size~$\approx$~0.37~pc) the expansion and the two points lie at the
lower end of the scatter of the linear correlation  for GPS/CSS
sources. This could be explained by projection effects. Since  III~Zw~2
is a Seyfert 1 galaxy with superluminal motion, the jet is probably
close to the line-of-sight. Hence the true size would be
underestimated, and the points in the plot move to the right.  However,
the evolution of III~Zw~2 during the expansion is almost parallel to
the correlation. This implies that the same physical processes,
i.e. synchrotron  self-absorption, are involved in III~Zw~2 and in
GPS/CSS sources. In the case of III~Zw~2, the radio source is much
older than the current outburst. Hence, it is possible that
some of the GPS/CSS sources are in fact not young, but only show
intermittent activity.

III~Zw~2 remains an extremely unusual object. Future simultaneous
multi-frequency observations of new outbursts would help to confirm the
proposed scenario of a jet-ISM interaction.

\begin{acknowledgements}
The National Radio Astronomy Observatory is a facility of the National
Science Foundation operated under cooperative agreement by Associated
Universities, Inc. The UMRAO is partially supported by funds from the
National Science Foundation and from the Univ. of Michigan Dept. of
Astronomy. The 100 m telescope at Effelsberg is operated by the
Max-Planck-Institut f\"ur Radioastronomie in Bonn. 
\end{acknowledgements}

\bibliography{brunthal_refs}

\begin{thebibliography}{}

\bibitem[\protect\citeauthoryear{{Aller} et~al.}{{Aller}
  et~al.}{1985}]{AllerAllerLatimer1985}
{Aller} H.~D., {Aller} M.~F., {Latimer} G.~E.,  {Hodge} P.~E., 1985, \apjs, 59,
  513

\bibitem[\protect\citeauthoryear{{Alonso-Herrero}, {Ward}, \&
  {Kotilainen}}{{Alonso-Herrero}
  et~al.}{1997}]{Alonso-HerreroWardKotilainen1997}
{Alonso-Herrero} A., {Ward} M.~J.,  {Kotilainen} J.~K., 1997, \mnras, 288, 977

\bibitem[\protect\citeauthoryear{{Arp}}{{Arp}}{1968}]{Arp1968}
{Arp} H., 1968, \apj, 152, 1101

\bibitem[\protect\citeauthoryear{{Bahcall}, {Kirhakos}, \&
  {Schneider}}{{Bahcall} et~al.}{1995}]{BahcallKirhakosSchneider1995}
{Bahcall} J.~N., {Kirhakos} S.,  {Schneider} D.~P., 1995, \apj, 450, 486

\bibitem[\protect\citeauthoryear{{Blandford} \& {Konigl}}{{Blandford} \&
  {Konigl}}{1979}]{BlandfordKonigl1979}
{Blandford} R.~D.,  {Konigl} A., 1979, \apj, 232, 34

\bibitem[\protect\citeauthoryear{{Blundell} \& {Beasley}}{{Blundell} \&
  {Beasley}}{1998}]{BlundellBeasley1998}
{Blundell} K.~M.,  {Beasley} A.~J., 1998, \mnras, 299, 165

\bibitem[\protect\citeauthoryear{{Blundell}, {Beasley}, \&
  {Bicknell}}{{Blundell} et~al.}{2003}]{BlundellBeasleyBicknell2003}
{Blundell} K.~M., {Beasley} A.~J.,  {Bicknell} G.~V., 2003, \apjl, 591, L103

\bibitem[\protect\citeauthoryear{{Boroson} \& {Green}}{{Boroson} \&
  {Green}}{1992}]{BorosonGreen1992}
{Boroson} T.~A.,  {Green} R.~F., 1992, \apjs, 80, 109

\bibitem[\protect\citeauthoryear{{Brunthaler} et~al.}{{Brunthaler}
  et~al.}{2000}]{BrunthalerFalckeBower2000}
{Brunthaler} A., {Falcke} H., {Bower} G.~C., et~al., 2000, \aap, 357, L45

\bibitem[\protect\citeauthoryear{{Clements} et~al.}{{Clements}
  et~al.}{1995}]{ClementsSmithAller1995}
{Clements} S.~D., {Smith} A.~G., {Aller} H.~D.,  {Aller} M.~F., 1995, \aj, 110,
  529

\bibitem[\protect\citeauthoryear{{Falcke} \& {Biermann}}{{Falcke} \&
  {Biermann}}{1995}]{FalckeBiermann1995}
{Falcke} H.,  {Biermann} P.~L., 1995, \aap, 293, 665

\bibitem[\protect\citeauthoryear{{Falcke} et~al.}{{Falcke}
  et~al.}{1999}]{FalckeBowerLobanov1999}
{Falcke} H., {Bower} G.~C., {Lobanov} A.~P., et~al., 1999, \apjl, 514, L17

\bibitem[\protect\citeauthoryear{{Falcke} et~al.}{{Falcke}
  et~al.}{2000}]{FalckeHenkelPeck2000}
{Falcke} H., {Henkel} C., {Peck} A.~B., et~al., 2000, \aap, 358, L17

\bibitem[\protect\citeauthoryear{{Falcke}, {Patnaik}, \& {Sherwood}}{{Falcke}
  et~al.}{1996}]{FalckePatnaikSherwood1996}
{Falcke} H., {Patnaik} A.~R.,  {Sherwood} W., 1996, \apjl, 473, L13

\bibitem[\protect\citeauthoryear{{Falcke}, {Sherwood}, \& {Patnaik}}{{Falcke}
  et~al.}{1996}]{FalckeSherwoodPatnaik1996}
{Falcke} H., {Sherwood} W.,  {Patnaik} A.~R., 1996, \apj, 471, 106

\bibitem[\protect\citeauthoryear{{Hutchings}}{{Hutchings}}{1983}]{Hutchings198%
3}
{Hutchings} J.~B., 1983, \pasp, 95, 799

\bibitem[\protect\citeauthoryear{{Hutchings} \& {Campbell}}{{Hutchings} \&
  {Campbell}}{1983}]{HutchingsCampbell1983}
{Hutchings} J.~B.,  {Campbell} B., 1983, \nat, 303, 584

\bibitem[\protect\citeauthoryear{{Kaastra} \& {de Korte}}{{Kaastra} \& {de
  Korte}}{1988}]{KaastradeKorte1988}
{Kaastra} J.~S.,  {de Korte} P.~A.~J., 1988, \aap, 198, 16

\bibitem[\protect\citeauthoryear{{Kellermann} et~al.}{{Kellermann}
  et~al.}{1989}]{KellermannSramekSchmidt1989}
{Kellermann} K.~I., {Sramek} R., {Schmidt} M., {Shaffer} D.~B.,  {Green} R.,
  1989, \aj, 98, 1195

\bibitem[\protect\citeauthoryear{{Kellermann} et~al.}{{Kellermann}
  et~al.}{1998}]{KellermannVermeulenZensus1998}
{Kellermann} K.~I., {Vermeulen} R.~C., {Zensus} J.~A.,  {Cohen} M.~H., 1998,
  \aj, 115, 1295

\bibitem[\protect\citeauthoryear{{Khachikian} \& {Weedman}}{{Khachikian} \&
  {Weedman}}{1974}]{KhachikianWeedman1974}
{Khachikian} E.~Y.,  {Weedman} D.~W., 1974, \apj, 192, 581

\bibitem[\protect\citeauthoryear{{Kirhakos} et~al.}{{Kirhakos}
  et~al.}{1999}]{KirhakosBahcallSchneider1999}
{Kirhakos} S., {Bahcall} J.~N., {Schneider} D.~P.,  {Kristian} J., 1999, \apj,
  520, 67

\bibitem[\protect\citeauthoryear{{Kukula} et~al.}{{Kukula}
  et~al.}{1998}]{KukulaDunlopHughes1998}
{Kukula} M.~J., {Dunlop} J.~S., {Hughes} D.~H.,  {Rawlings} S., 1998, \mnras,
  297, 366

\bibitem[\protect\citeauthoryear{{Lloyd}}{{Lloyd}}{1984}]{Lloyd1984}
{Lloyd} C., 1984, \mnras, 209, 697

\bibitem[\protect\citeauthoryear{{Miller}, {Rawlings}, \& {Saunders}}{{Miller}
  et~al.}{1993}]{MillerRawlingsSaunders1993}
{Miller} P., {Rawlings} S.,  {Saunders} R., 1993, \mnras, 263, 425

\bibitem[\protect\citeauthoryear{{O'Dea} \& {Baum}}{{O'Dea} \&
  {Baum}}{1997}]{ODeaBaum1997}
{O'Dea} C.~P.,  {Baum} S.~A., 1997, \aj, 113, 148

\bibitem[\protect\citeauthoryear{{Osterbrock}}{{Osterbrock}}{1977}]{Osterbrock%
1977}
{Osterbrock} D.~E., 1977, \apj, 215, 733

\bibitem[\protect\citeauthoryear{{Peck} et~al.}{{Peck}
  et~al.}{2003}]{PeckHenkelUlvestad2003}
{Peck} A.~B., {Henkel} C., {Ulvestad} J.~S., et~al., 2003, \apj, 590, 149

\bibitem[\protect\citeauthoryear{{Ryle} \& {Longair}}{{Ryle} \&
  {Longair}}{1967}]{RyleLongair1967}
{Ryle} M.~S.,  {Longair} M.~S., 1967, \mnras, 136, 123

\bibitem[\protect\citeauthoryear{{Salvi} et~al.}{{Salvi}
  et~al.}{2002}]{SalviPageStevens2002}
{Salvi} N.~J., {Page} M.~J., {Stevens} J.~A., et~al., 2002, \mnras, 335, 177

\bibitem[\protect\citeauthoryear{{Scarpa} et~al.}{{Scarpa}
  et~al.}{2000}]{ScarpaUrryFalomo2000}
{Scarpa} R., {Urry} C.~M., {Falomo} R., {Pesce} J.~E.,  {Treves} A., 2000,
  \apj, 532, 740

\bibitem[\protect\citeauthoryear{{Schmidt} \& {Green}}{{Schmidt} \&
  {Green}}{1983}]{SchmidtGreen1983}
{Schmidt} M.,  {Green} R.~F., 1983, \apj, 269, 352

\bibitem[\protect\citeauthoryear{{Shepherd}, {Pearson}, \& {Taylor}}{{Shepherd}
  et~al.}{1994}]{ShepherdPearsonTaylor1994}
{Shepherd} M.~C., {Pearson} T.~J.,  {Taylor} G., 1994, BAAS, 26, 987

\bibitem[\protect\citeauthoryear{{Surace}, {Sanders}, \& {Evans}}{{Surace}
  et~al.}{2001}]{SuraceSandersEvans2001}
{Surace} J.~A., {Sanders} D.~B.,  {Evans} A.~S., 2001, \aj, 122, 2791

\bibitem[\protect\citeauthoryear{{T{\" u}rler}, {Courvoisier}, \&
  {Paltani}}{{T{\" u}rler} et~al.}{1999}]{TuerlerCourvoisierPaltani1999}
{T{\" u}rler} M., {Courvoisier} T.~J.-L.,  {Paltani} S., 1999, \aap, 349, 45

\bibitem[\protect\citeauthoryear{{Taylor} et~al.}{{Taylor}
  et~al.}{1996}]{TaylorDunlopHughes1996}
{Taylor} G.~L., {Dunlop} J.~S., {Hughes} D.~H.,  {Robson} E.~I., 1996, \mnras,
  283, 930

\bibitem[\protect\citeauthoryear{{Unger} et~al.}{{Unger}
  et~al.}{1987}]{UngerLawrenceWilson1987}
{Unger} S.~W., {Lawrence} A., {Wilson} A.~S., {Elvis} M.,  {Wright} A.~E.,
  1987, \mnras, 228, 521

\bibitem[\protect\citeauthoryear{{Valtaoja} et~al.}{{Valtaoja}
  et~al.}{1999}]{ValtaojaLaehteenmaekiTeraesranta1999}
{Valtaoja} E., {L{\" a}hteenm{\" a}ki} A., {Ter{\" a}sranta} H.,  {Lainela} M.,
  1999, \apjs, 120, 95

\bibitem[\protect\citeauthoryear{{Zwicky}}{{Zwicky}}{1967}]{Zwicky1967}
{Zwicky} F., 1967, Adv. Astron. Astrophys., 5, 267

\end{thebibliography}
\bibliographystyle{aa}

\end{document}